\documentclass[tighten,twocolumn]{aastex62}

\usepackage{color}

\usepackage{graphicx}
\usepackage{rotating}
\usepackage{float}

\def\kms{$\mathrm{km \,s}^{-1}$}

\usepackage{enumitem}

\begin{document}

\shorttitle{Continuum enhancements, evolution of line profiles and magnetic field during flares}
\shortauthors{Zuccarello et al.}

\title{Continuum enhancements, line profiles and magnetic field evolution during consecutive flares}

\author{Francesca Zuccarello}
\affiliation{Dipartimento di Fisica e Astronomia ``Ettore Majorana'' - Sezione Astrofisica, Universit\`{a} degli Studi di Catania,
	Via S.~Sofia 78, I-95123 Catania, Italy}

\author{Salvo L. Guglielmino}
\affiliation{Dipartimento di Fisica e Astronomia ``Ettore Majorana'' - Sezione Astrofisica, Universit\`{a} degli Studi di Catania,
	Via S.~Sofia 78, I-95123 Catania, Italy}

\author{Vincenzo Capparelli}
\affiliation{Dipartimento di Fisica e Astronomia ``Ettore Majorana'' - Sezione Astrofisica, Universit\`{a} degli Studi di Catania,
	Via S.~Sofia 78, I-95123 Catania, Italy}

\author{Mihalis Mathioudakis}
\affiliation{Astrophysics Research Centre, School of Mathematics \& Physics, Queen's University Belfast, Belfast, BT7 1NN, UK}

\author{Peter H. Keys}
\affiliation{Astrophysics Research Centre, School of Mathematics \& Physics, Queen's University Belfast, Belfast, BT7 1NN, UK}

\author{Serena Criscuoli}
\affiliation{NSO -- National Solar Observatory, Sacramento Peak - Box 62, Sunspot NM 88349, USA}

\author{Mariachiara Falco}
\affiliation{INAF - Osservatorio Astrofisico di Catania,
	Via S.~Sofia 78, I-95123 Catania, Italy}

\author{Mariarita Murabito}
\altaffiliation{currently at INAF - Osservatorio Astronomico di Roma, \\
	via Frascati 33, I-00078 Monte Porzio Catone, Italy}
\affiliation{Dipartimento di Fisica e Astronomia ``Ettore Majorana'' - Sezione Astrofisica, Universit\`{a} degli Studi di Catania,
	Via S.~Sofia 78, I-95123 Catania, Italy}

\correspondingauthor{Salvo~L. Guglielmino}
\email{salvatore.gugliemino@inaf.it}

\begin{abstract}

During solar flares, magnetic energy can be converted into electromagnetic radiation from radio waves to $\gamma$ rays. Enhancements in the continuum at visible wavelengths give rise to  white-light flares, as well as continuum enhancements in the FUV and NUV passbands. In addition, the strong energy release in these events can lead to the rearrangement of the magnetic field at the photospheric level, causing morphological changes in large and stable magnetic structures like sunspots.
In this context, we describe observations acquired by satellite instruments (\textit{IRIS}, \textit{SDO}/HMI, \textit{Hinode}/SOT) and ground-based telescopes (ROSA/DST) during two consecutive C7.0 and X1.6 flares occurred in active region NOAA~12205 on 2014 November~7. The flare was accompanied by an eruption. The results of the analysis show the presence of continuum enhancements during the evolution of the events, observed both in ROSA images and in \textit{IRIS} spectra. In the latter, a prominent blue-shifted component is observed at the onset of the eruption. We investigate the role played by the evolution of the $\delta$ sunspots of the active region in the flare triggering, and finally we discuss the changes in the penumbrae surrounding these sunspots as a further consequence of these flares.

\end{abstract}



\section{Introduction}

Solar flares are complex eruptive phenomena occurring in the atmospheric layers of the Sun, releasing energy spanning typically from 10$^{28}$ to 10$^{32}$~erg. They are often triggered by the destabilization of a filament located above a polarity inversion line (PIL; \citealp[see, e.g.,][for a review and references therein]{fletcher}). This energy, previously stored in a non-potential magnetic field configuration, is converted, through magnetic reconnection, in kinetic energy, bulk plasma motions and electromagnetic radiation emitted through the whole spectrum, from decameter radio waves to gamma rays at 100~MeV \citep[see, e.g.,][]{benz}. Far-reaching consequences may be due to such strong energy release phenomena, with potential impact on the Earth \citep{Zuccarello:13,Patsourakos:16,Piersanti:17}.

Most of the flare emission signatures observed at different wavelengths can be explained in the framework of the CSHKP two-dimensional (2D) magnetic reconnection model, named for \citet{carmi}, \citet{sturr}, \citet{hira}, and \citet{kopp}: in a coronal arcade a magnetic reconnection process caused, for instance, by the eruption of a filament, takes place, so that the field lines of the arcade confining the flux rope, get opened and later reconnect. At the increasing reconnection heights that characterize the arcade field lines, from the most internal to the most external, electrons and protons are accelerated and, when colliding with the lower atmospheric layers, can impulsively heat the local plasma and cause the chromospheric evaporation which fills the post-flare loops \citep{Milligan:15}. 

A more realistic three-dimensional (3D) treatment of magnetic reconnection allows the interpretation of some of the flare properties that cannot be understood in the framework of the CSHKP model, such as the evolution of the shear of flare loops, their morphology and relative positioning, and the motions of EUV or X-ray sources along the ribbons (\citealp[e.g.][]{Aulanier:12,Aulanier:13,Janvier:13}; more recently, \citealp{Janvier:17}). The 3D reconnection occurs in the quasi-separatrix layers \citep[e.g.][and references therein]{Demoulin:06}, where the displacement of the magnetic field lines determines a continuous exchange of connectivities with the neighboring field lines. In the lower atmosphere, this process appears as an apparent slipping motion of the field-line footpoints, which correspond to the flare ribbons and is referred to as \textit{slipping reconnection} \citep{Aulanier:06,Aulanier:07}. This 3D solar flare model has been verified in many observations \citep{Dudik:14,Dudik:16,Sobotka:16,Zheng:16}. In particular, such complex dynamics seem to be favoured when flares occur in active regions with intricate magnetic configurations, like those hosting $\delta$ sunspots, where a fan-spine topology is usually formed \citep[e.g.,][]{Guglielmino:16}.


\begin{figure*}[t]
	\centering
	\includegraphics[trim=20 20 80 10, clip, scale=0.325]{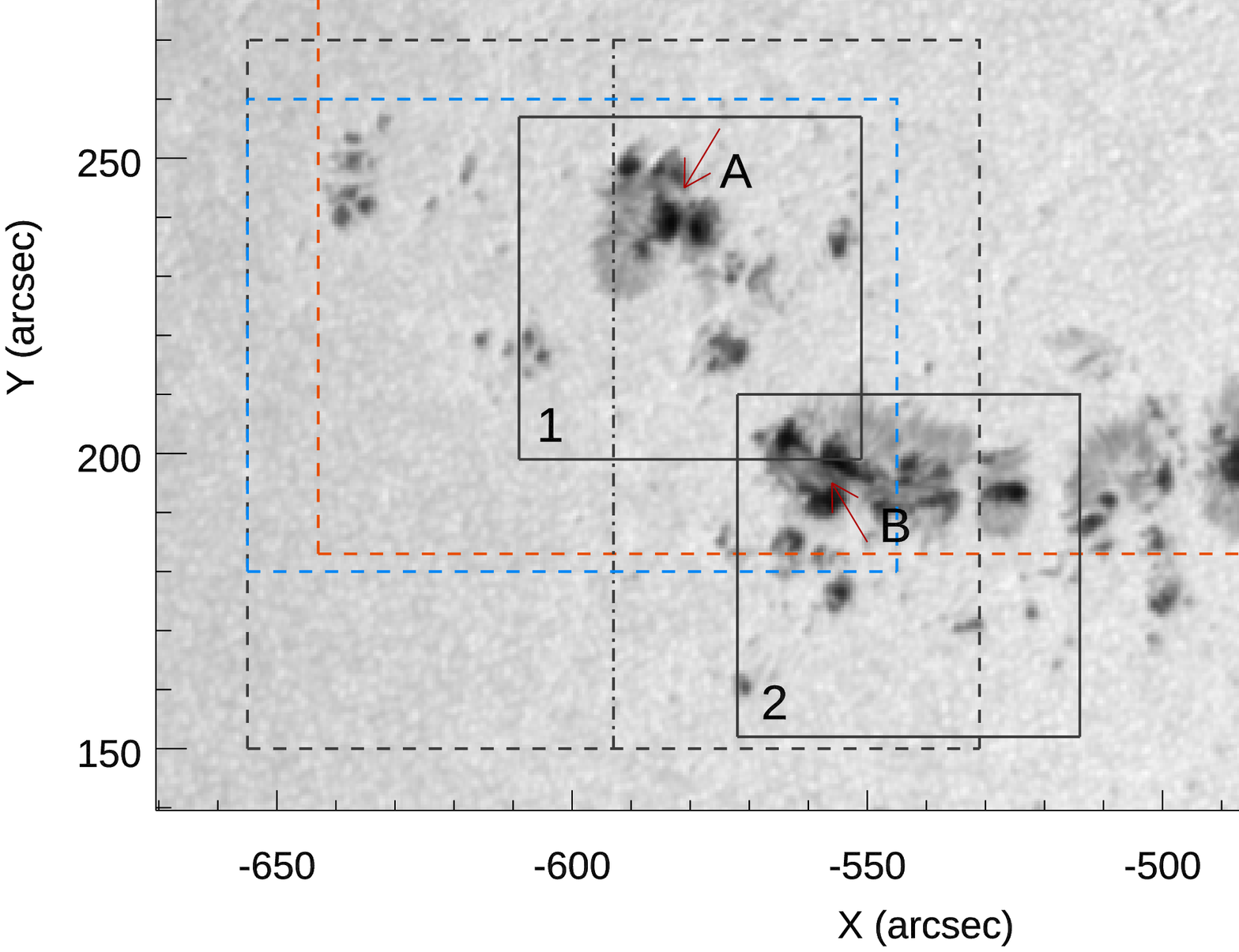}%
	\hspace*{2ex}
	\includegraphics[trim=80 20 80 10, clip, scale=0.325]{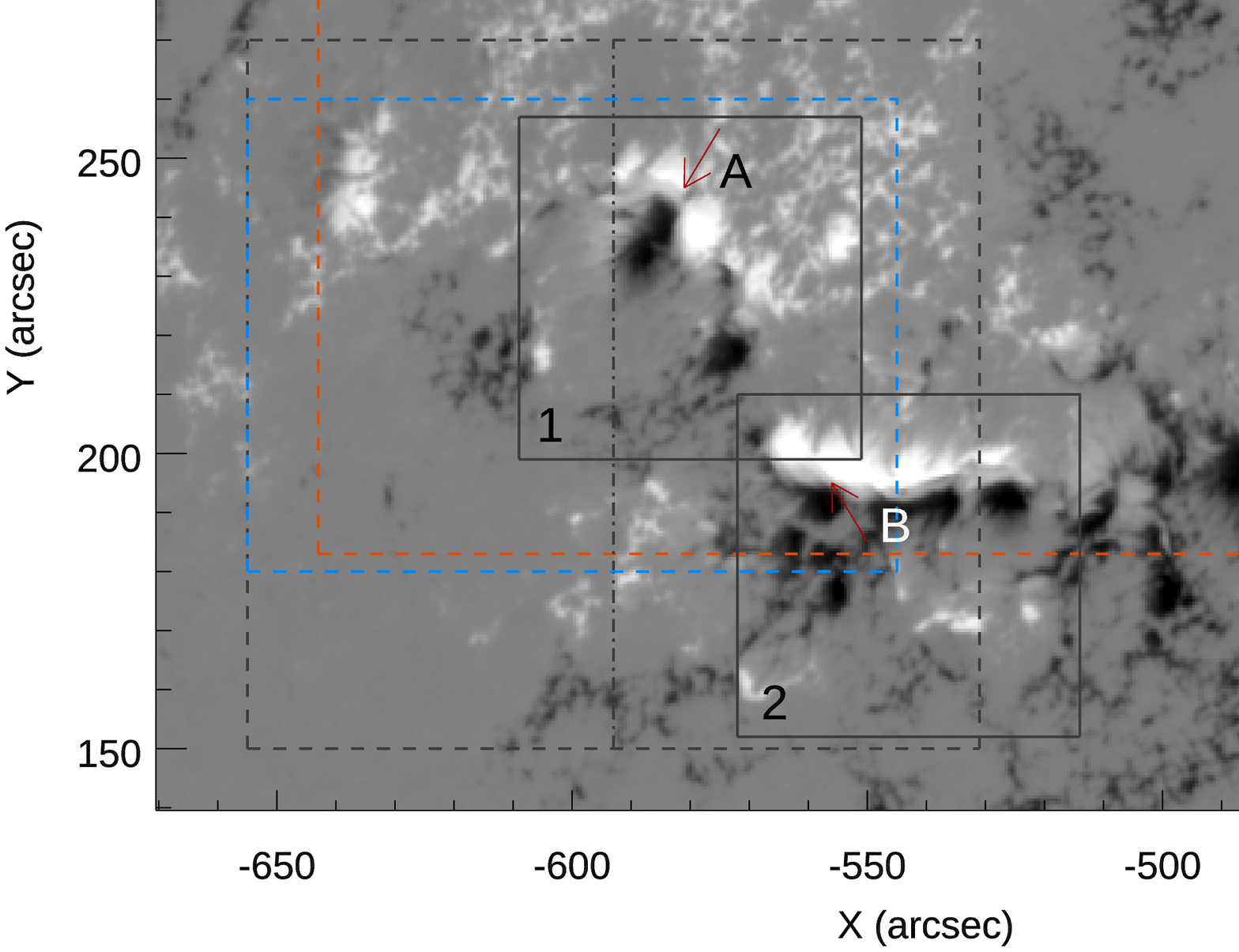}
	\caption{\textit{Left}: \textit{SDO}/HMI continuum intensity map showing the morphology of AR~12205 on 2014 November~7 at 16:58:12~UT. \textit{Right}: \textit{SDO}/HMI line-of-sight magnetogram showing the magnetic configuration of the AR at the same time. In both maps the dashed black box indicates the \textit{IRIS} SJI FOV and the dotted-dashed vertical line shows the approximate position of the slit during the \textit{IRIS} raster observations; the dashed blue box indicates the \textit{IRIS} SJI sub-FOV used in Figure~\ref{fig:iris_fov}. The two solid-line boxes indicate the FOVs acquired by the ROSA instrument at different times. Labels 1 and 2 in these boxes indicate the FOV1 and FOV2, respectively (see main text). The red dashed box indicates the \textit{Hinode}/SOT FOV. The red arrows in both maps indicate the presence of two $\delta$ spots, labelled with letters \textit{A} and \textit{B}, respectively. In these and in the following images, if not otherwise specified, north is on the top, west is to the right.} \label{fig:sdo_context}
\end{figure*}


In the visible range, the emission is commonly observed in the H$\alpha$ line and is believed to be due to the effect of collisions on the chromospheric plasma by the energetic particles accelerated at the reconnection site and precipitating towards lower heights. This emission, which displays an increase already during the impulsive phase, is often observed in the form of two bright ribbons, located parallel to the PIL and separating from each other with an initial velocity of $\sim 100$~\kms, decreasing to less than $1$~\kms{} in the following tens of minutes/hours \citep[see, e.g.,][]{Maurya:09}.
Indeed, taking into account the 3D nature of magnetic reconnection, in the presence of a fan-spine topology, flares may exhibit extra ribbons or even circular ribbons \citep{Masson:09,Romano:17}. 

Observations show that increased emission is also detected in the visible continuum, i.e., in white light (WL), and is often spatially and temporally correlated with the hard X-ray (HXR, $\sim 10$~keV) signatures \citep[see, e.g.,][]{hudson}. In this regard, we recall that WL flares were initially believed to be associated only with large X-class events, that is when the EUV or soft X-ray emission could exceed a certain threshold \citep{neidig83}. However, successive observations provided evidence of the presence in the continuum of bright kernels of size $\approx 1\arcsec - 3\arcsec$, whose location was cospatial with the more extended flare ribbons detected in the core of strong lines, like the H$\alpha$ and \ion{Ca}{2} lines \citep{neidig89}. More recently, \citet{jess08} have shown that WL emission can be observed also during C-class flares. 

As a result of the rapid restructuring of the 3D magnetic field driven by magnetic reconnection, photospheric structures can be affected by flares. Sudden and irreversible changes of the photospheric magnetic field may occur in spite of the large inertia of the photosphere \citep[see the reviews of][]{WangLiu:15,Toriumi:19}. In particular, considering also WL observations, it has been observed that penumbral structures usually decay in the peripheral sides of $\delta$ sunspots hosting flares, whereas they darken near the flaring PILs. Several recent observations demonstrated that the transverse field is enhanced at the central flaring PILs \citep[e.g.,][]{Petrie:10,Wang:12a,Wang:12b,Petrie:12,Petrie:13,Wang:14}. In general, most of the changes are localized around the PIL regions \citep{Castellanos:18,Lu:19}.

In this framework, the main goal of this work is to investigate if and how continuum emission and changes in line profiles during flares are associated to rearrangements in the magnetic field. We study two consecutive flares that occurred in active region (AR) NOAA~12205 (hereafter, AR~12205) on 2014 November~7 using data acquired from ground-based and satellite instruments. The strongest flare, classified as X1.6 event, was also analyzed by \citet{yurchy15} using different datasets. We identify the complex configuration of AR~12205, comprising two $\delta$ sunspots, as the possible flare trigger. We study the characteristics of the emission both in UV spectral lines and in the continuum (UV and visible wavelengths) during the evolution of the events. We also correlate the observed continuum changes induced in the photospheric structures to the rearrangement of the magnetic field. 

The paper is organized as follows: in Sect.~2 we describe the observational data. In Sect.~3 the data analysis and the results are reported. In Sect.~4 we discuss the results, describing our conclusions in Sect.~5.

\section{Observations}

AR~12205, located at N15E33 on 2014 November~7, was characterized by a $\beta \gamma \delta$ magnetic configuration and produced several flares with various X-ray classifications. In particular, on 2014 November~7, an M1.0 flare (peak at 10:13~UT) and a series of C-class flares (C3.9, peak at 12:03~UT; C1.3, 13:19~UT; C2.3, 13:55~UT; C7.0, 14:51~UT; C7.0, 16:10~UT) occurred before the onset of an X1.6 flare (peak at 17:26~UT).

\begin{figure}[!b] 
	\centering
	\includegraphics[trim=35 0 2 0, clip,scale=.30]{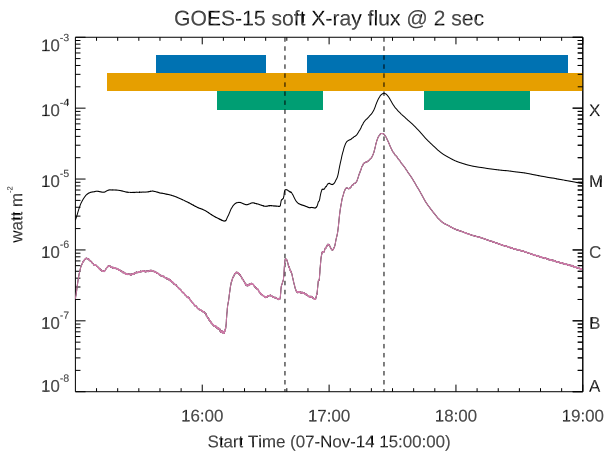} 
	\caption{GOES soft X-rays in two wavelength ranges: $1.0 - 8.0$~\AA{} (black line) and $0.5 - 4$~\AA{} (magenta line) showing the flux increase during the flares analyzed in the paper. The peak time of the C7.0 (16:39~UT) and X1.6 (17:26~UT) flares are indicated by vertical dashed lines. The blue rectangles indicate the time intervals of acquisition of the ROSA instrument for FOV1 (left) and FOV2 (right). Analogously, the orange rectangle indicates the time interval of acquisition of the \textit{Hinode}/SOT, and the green rectangles indicate the time intervals of acquisition of the \textit{IRIS} instrument.} \label{fig:goes}
\end{figure}

We observed this active region during a coordinated observing campaign carried out using the ground-based Rapid Oscillations in the Solar Atmosphere \citep[ROSA;][]{jess10} imaging system mounted at the Dunn Solar Telescope (DST) at the US National Solar Observatory in New Mexico and the Interface Region Imaging Spectrograph \citep[\textit{IRIS};][]{depontieu} satellite (see Figures~\ref{fig:sdo_context} and~\ref{fig:goes}). 

\begin{figure*}[t] 
	\centering
	\includegraphics[trim=70 20 120 50,clip,width=0.475\textwidth]{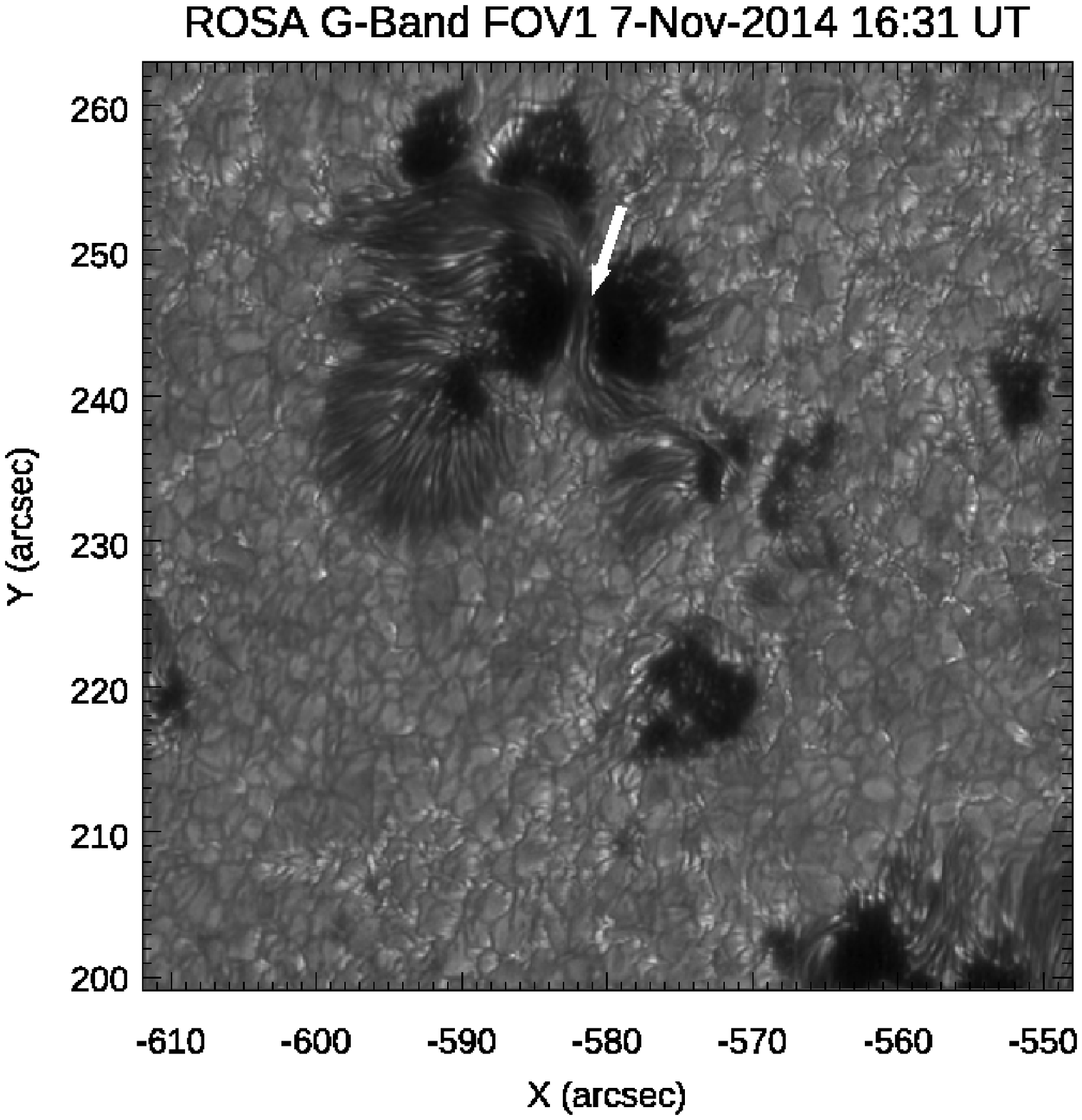}%
	\includegraphics[trim=70 20 120 50,clip,width=0.475\textwidth]{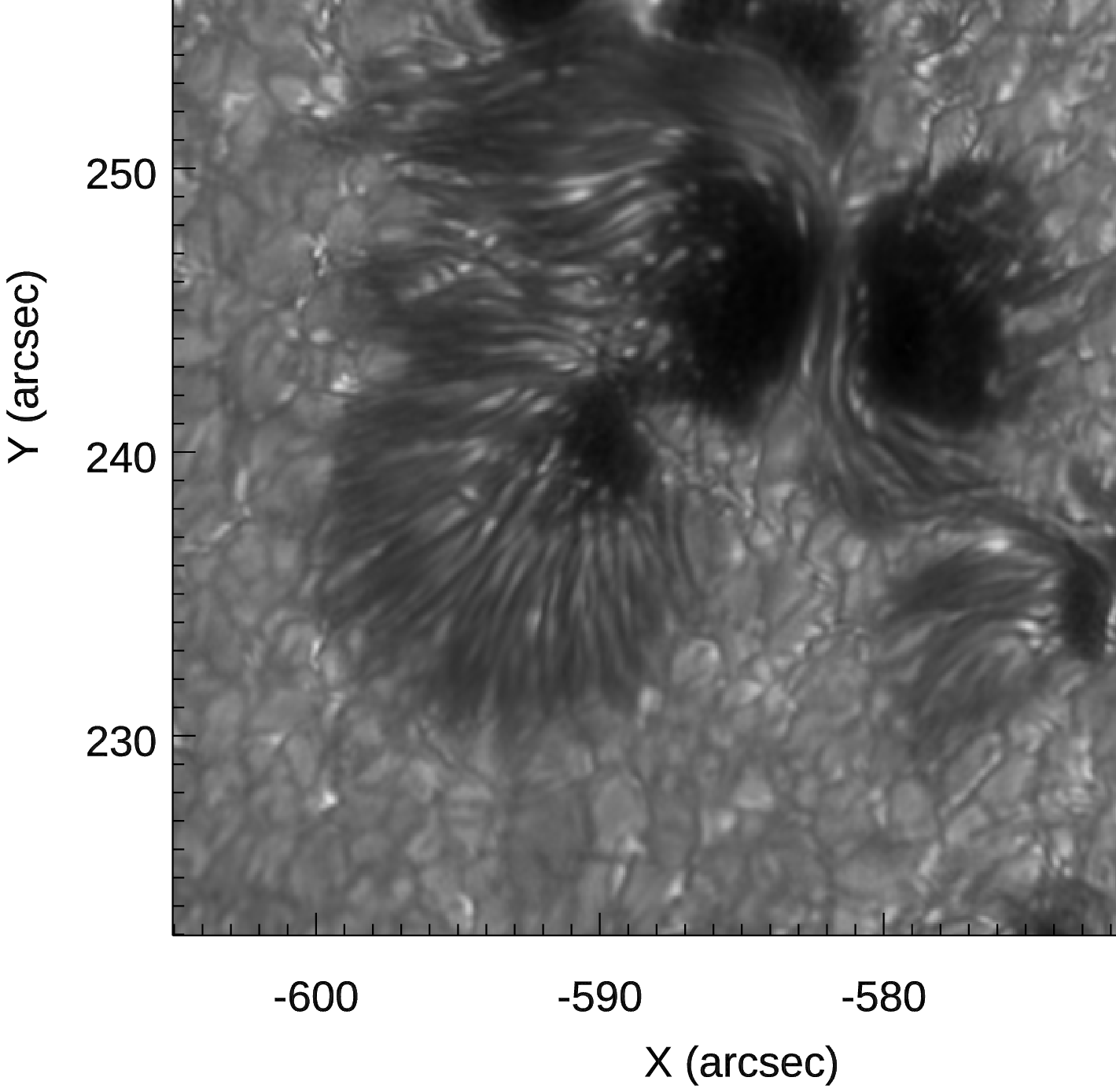}
	\hspace*{-0.1cm}\\
	\includegraphics[trim=70 20 120 50,clip,width=0.475\textwidth]{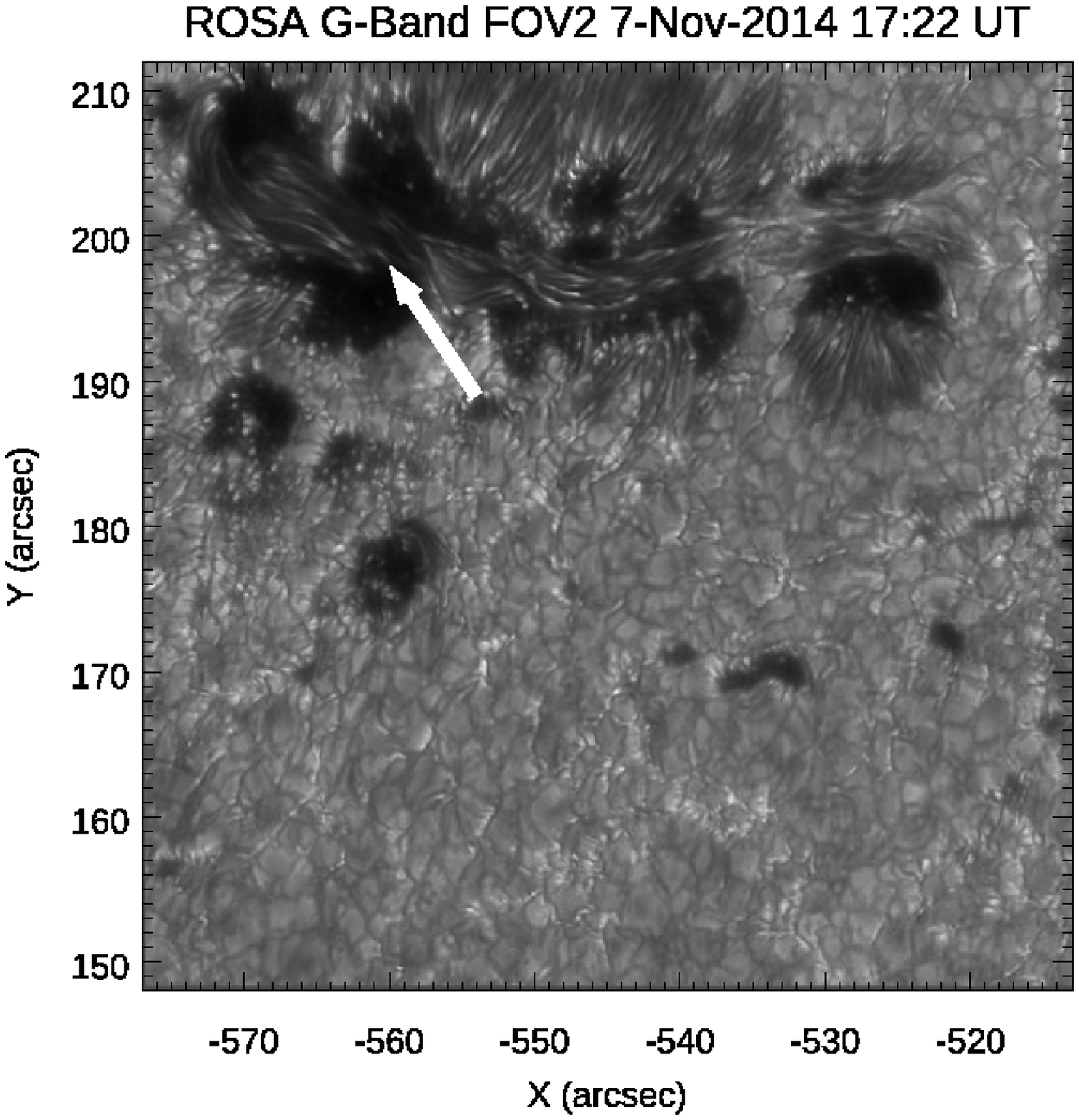}%
	\includegraphics[trim=70 20 120 50,clip,width=0.475\textwidth]{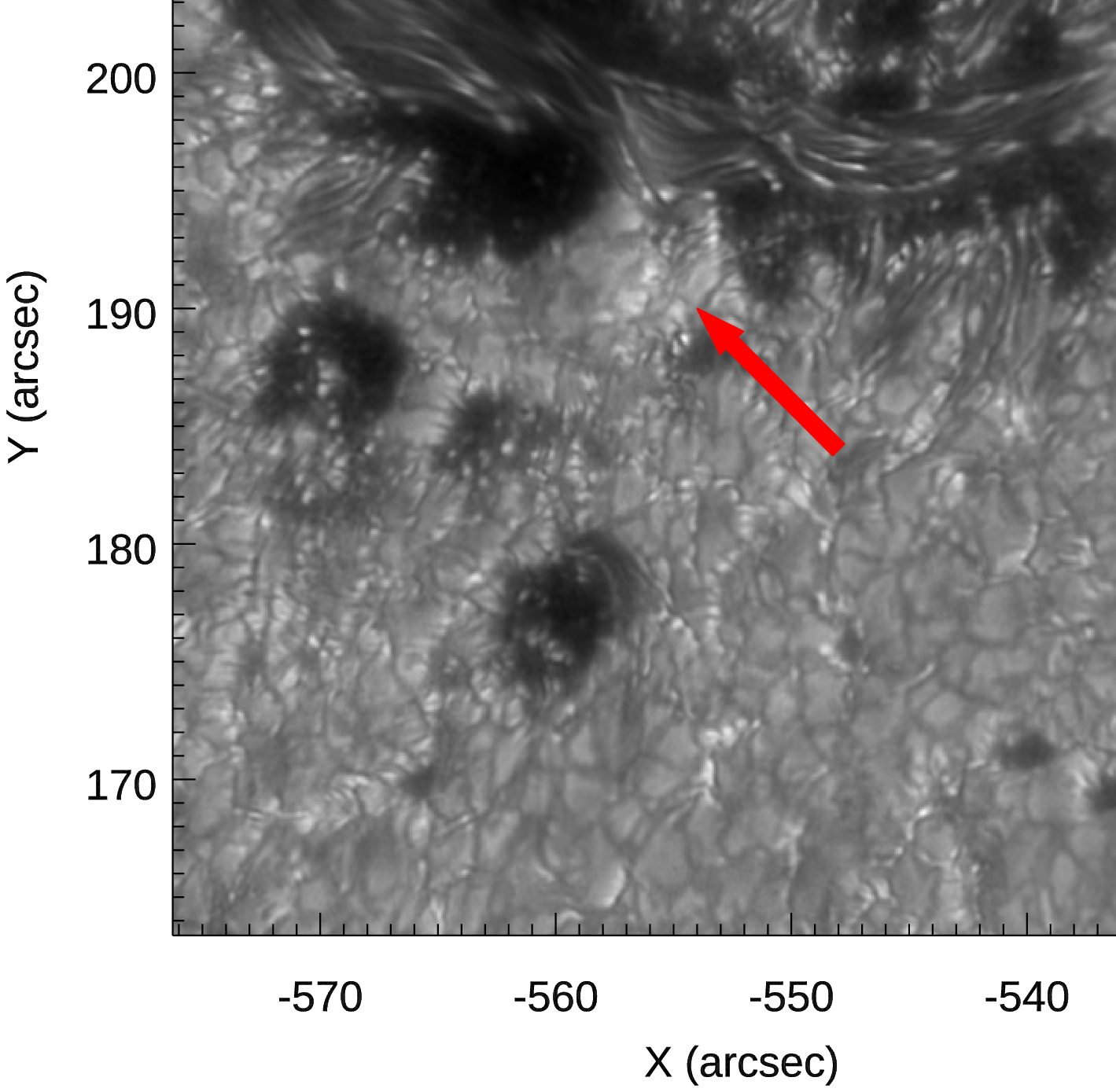}
	\hspace*{-0.1cm}\\
	\includegraphics[trim=70 20 120 50,clip,width=0.475\textwidth]{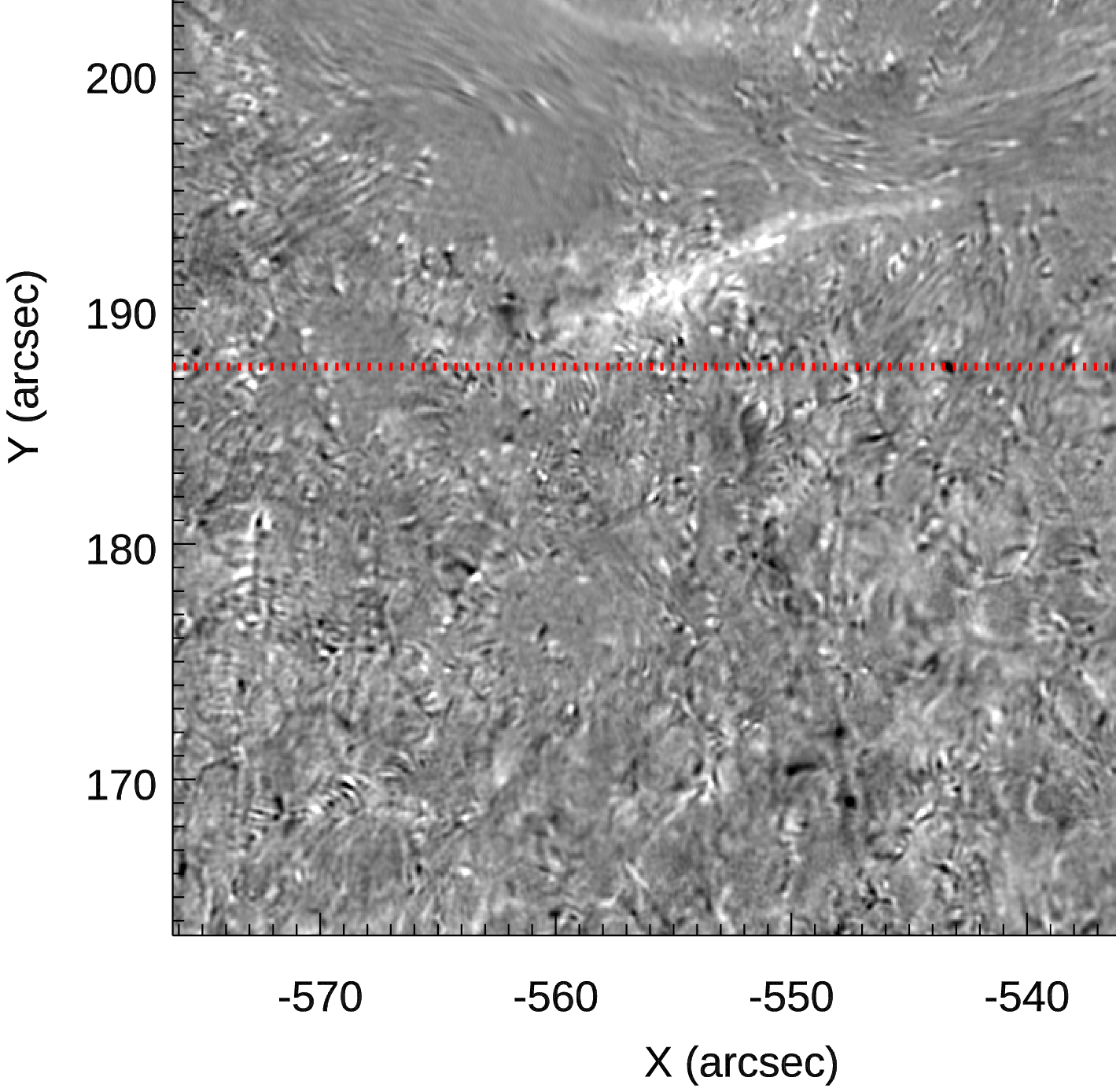}%
	\includegraphics[trim=70 20 120 50,clip,width=0.475\textwidth]{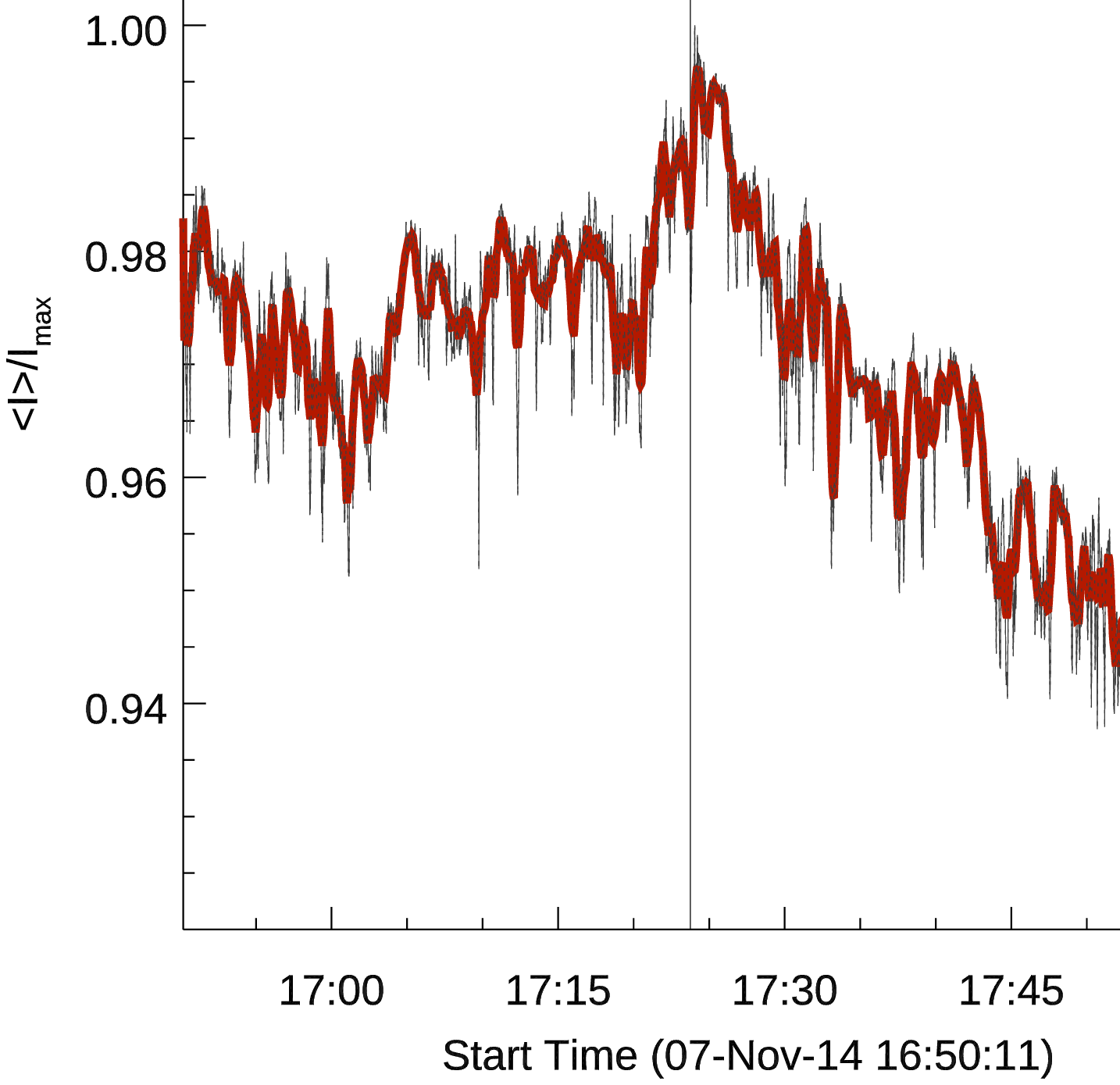}	 
	\caption{ROSA \textit{G}-band images of AR~12205. \textit{Top panels}: \textit{Left}: The entire FOV1, at 16:31~UT. The white arrow indicates the $\delta$-spot \textit{A}. \textit{Right}: Zoomed image from ROSA FOV1 showing the details of $\delta$-spot \textit{A}, characterized by the presence of sheared penumbral filaments within the two opposite magnetic polarities. \textit{Middle} panels: \textit{Left}: The entire FOV2 at 17:22~UT, during the rise phase of the X1.6 flare, a few minutes before the peak. The white arrow indicates the $\delta$-spot \textit{B}. \textit{Right}: Zoomed image from ROSA FOV2 showing the location of a ribbon observed in the continuum (red arrow) at 17:22~UT. \textit{Bottom panels}: \textit{Left}: Difference imaging of zoomed FOV2, showing the ribbons at a time close to the peak of the X1.6 flare. The red dotted line separates the zoomed FOV2 into two halves. \textit{Right}: \textit{G}-band lightcurve relevant to the upper half of zoomed FOV2, where the ribbons are observed. The red line represents a smoothed trend. The vertical line indicates the X1.6 flare peak.\\
	(An animation of the \textit{G}-band images relevant to zoomed FOV2 is available in the online material.)
	\label{fig:rosa_fov}}
\end{figure*}

The ROSA data were acquired simultaneously in the \ion{Ca}{2}~K core at 3933.7~\AA{} (bandpass 1.0~\AA), in the Blue continuum at 4170~\AA{} (bandpass 52.0~\AA) and in the \textit{G} band at 4305.5~\AA{} (bandpass 9.2~\AA), in two different fields-of-view (FOVs): FOV1 was acquired between 15:38 and 16:31~UT, while FOV2 was acquired between 16:50 and 18:53~UT (see the solid-line squares in Figure~\ref{fig:sdo_context} and the blue rectangles in Figure~\ref{fig:goes}). The \ion{Ca}{2}~K, \textit{G}-band and continuum observations were obtained with a diffraction limited spatial sampling of $0\farcs069$/pixel. The total FOV is $69\arcsec \times 69\arcsec$. High-order adaptive optics were applied throughout the observations \citep{rim}. The images were then reconstructed by implementing the speckle algorithms of \cite{wog} followed by de-stretching. These algorithms removed the effects of atmospheric distortion from the data. The effective cadence after reconstruction is reduced to 2.3~s for \ion{Ca}{2}~K and 2.112~s for the \textit{G} band and continuum. Moreover, during the speckle reconstruction, an apodisation windowing function is applied to the images to reduce artefacts introduced by Fourier transforms. This process reduces the FOV of the images to  $58\farcs65 \times 58\farcs65$ (see Figure~\ref{fig:rosa_fov}). 

\textit{IRIS} acquired SJI images in 3 passbands (\ion{C}{2} at 1330~\AA, \ion{Mg}{2}~k at 2796~\AA, and \ion{Mg}{2} wing at 2832~\AA). The large 4-step coarse raster mode was used for the slit, acquiring data from 16:07 to 16:57~UT and from 17:45 to 18:35~UT (see the green rectangles in the upper part of the plot shown in Figure~\ref{fig:goes}). The SJI filtergrams (80 for each interval of acquisition) were characterized by a FOV of $119\arcsec \times 119\arcsec$ with a sampling of $0\farcs166$ pixel$^{-1}$ and a temporal cadence of 37~s. The FOV of the raster was $6\arcsec \times 119\arcsec$ with a sampling of $0\farcs166$ pixel$^{-1}$ and a temporal cadence of 37~s, with a step cadence of 9.4~s. 

The analysis carried out in this study is based mainly on the \ion{Mg}{2}~k \&~h lines at 2796.35~\AA{} and 2803.55~\AA{} (characterized by a formation temperature of $\log \mathrm{T}$~[K] $= 4.0$), as well as \ion{C}{2} at 1330~\AA{} and \ion{Si}{4} at 1403~\AA{} (having formation temperatures of $\log \mathrm{T}$~[K] $= 4.3$ and~$4.8$, respectively) \citep{depontieu}. The central wavelength for each analyzed line has been determined by ensuring that the cool \ion{S}{1} 1401.515~\AA{} and \ion{Ni}{2} 2799.474~\AA{} lines were at rest within $\pm 5$~\kms{} for the FUV and NUV channels, respectively.

\begin{figure*}
	\centering
	\includegraphics[trim=0 40 0 30, clip, scale=.6]{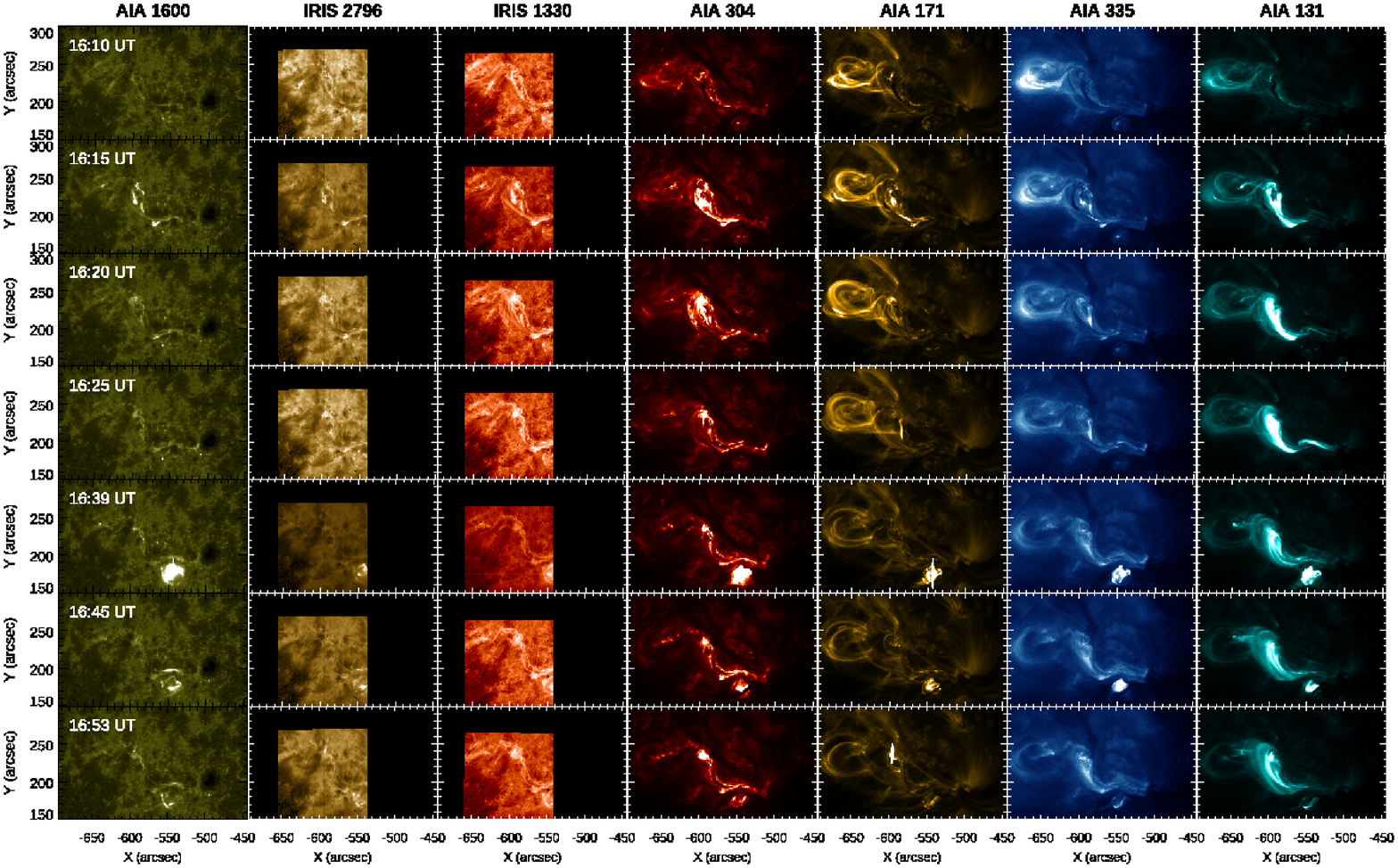}		
	\caption{Synoptic view of the evolution of the C7.0 flare at different atmospheric layers, at representative times during the observing interval. From left to right, each row shows the \textit{SDO}/AIA 1600~\AA{} map, the \textit{IRIS} SJI 2796~\AA{} and 1330~\AA{} maps, and the \textit{SDO}/AIA cospatial maps for the selected EUV channels. \label{fig:flareC}}
\end{figure*}

Context images acquired by the Helioseismic and Magnetic Imager \citep[HMI;][]{Scherrer:12} instrument onboard the \textit{Solar Dynamics Observatory} \citep[\textit{SDO},][]{Pesnell:12} satellite were used to obtain  information on the global magnetic field configuration of AR~12205. More precisely, full-disk continuum images and longitudinal (line-of-sight, LOS) magnetograms taken by HMI in the \ion{Fe}{1} line at 6173~\AA{} with a resolution of 1\arcsec{} were used to complement the high-resolution dataset of the ground-based instruments. In the present work, we also considered full-disk data from the Atmospheric Imaging Assembly \citep[AIA;][]{Lemen:12} on board the \textit{SDO} satellite. We used images from the 1600, 304, 171, 335, and 131~\AA{} channels. The cadence of the \textit{SDO}/AIA data is 24~s for the UV channel and 12~s for the EUV channels, respectively, with a spatial scale of about 0\farcs6 pixel$^{-1}$. \textit{SDO}/AIA images were also compensated for solar rotation effects.

We also benefit from high-resolution photospheric observations obtained with  the Solar Optical Telescope \citep[SOT,][]{Tsuneta:08} aboard the \textit{Hinode} satellite \citep{Kosugi:07}. A sequence of filtergrams in the \textit{G} band ($430.5 \pm 0.8$~nm) on AR~12205 was acquired between 15:15~UT and 19:06~UT on November~7, with an uneven cadence of about 10~min. A simultaneous sequence was acquired in the \ion{Ca}{2} H line ($396.85 \pm 0.3$~nm), with a cadence of 1~min. These data cover a FOV of about $223\arcsec \times 111\farcs5$.

\begin{figure}
	\centering
	\includegraphics[trim=20 265 60 230, clip, scale=.39]{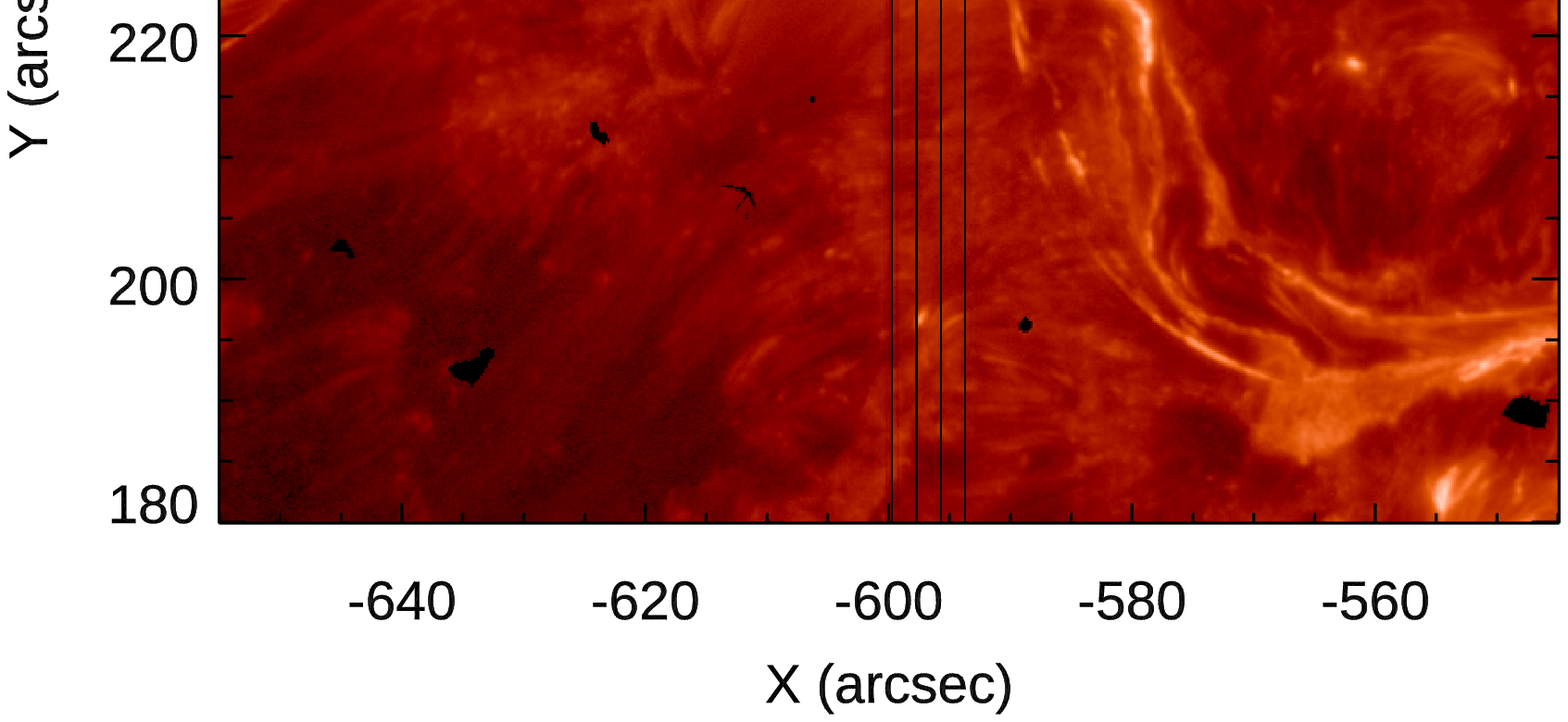}		
	\includegraphics[trim=20 265 60 230, clip, scale=.39]{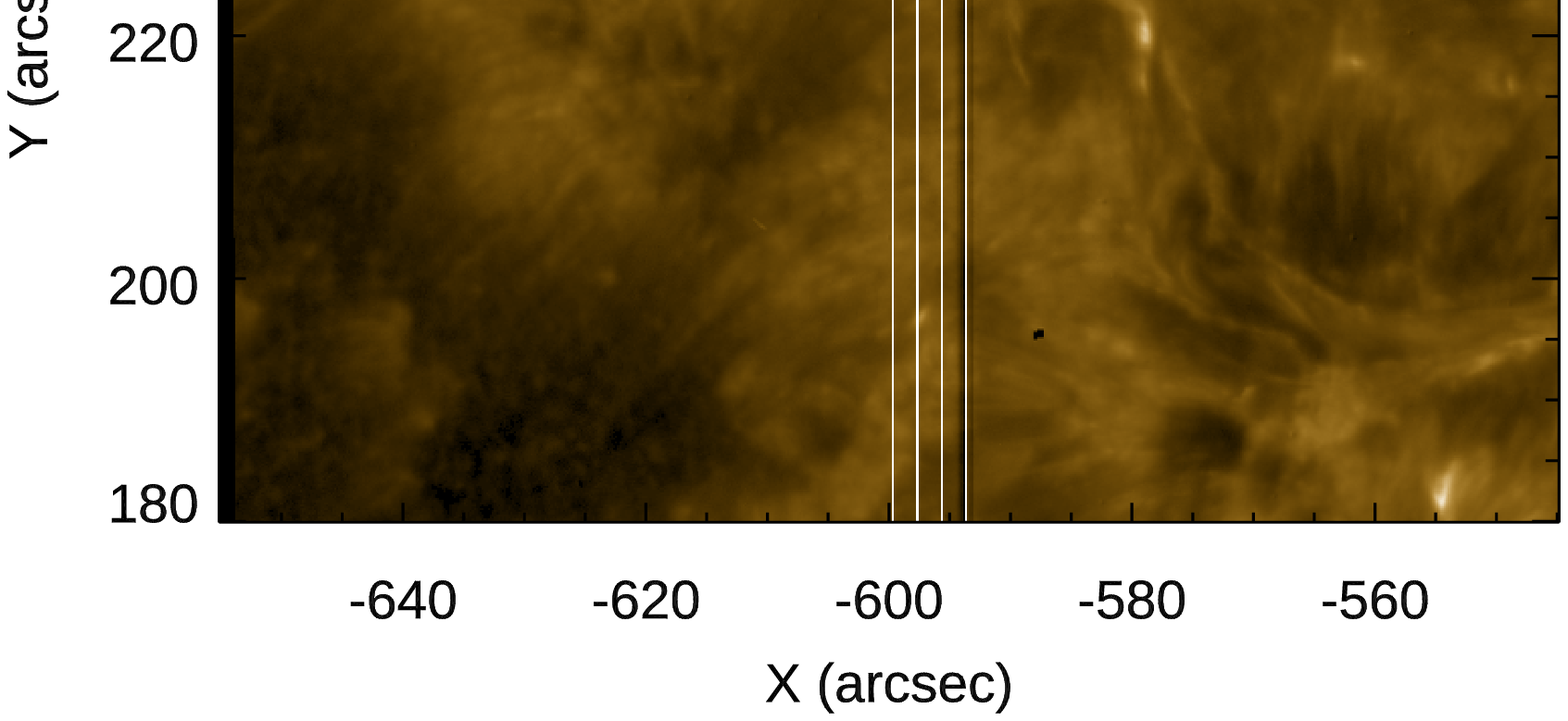}
	\includegraphics[trim=20 265 60 230, clip, scale=.39]{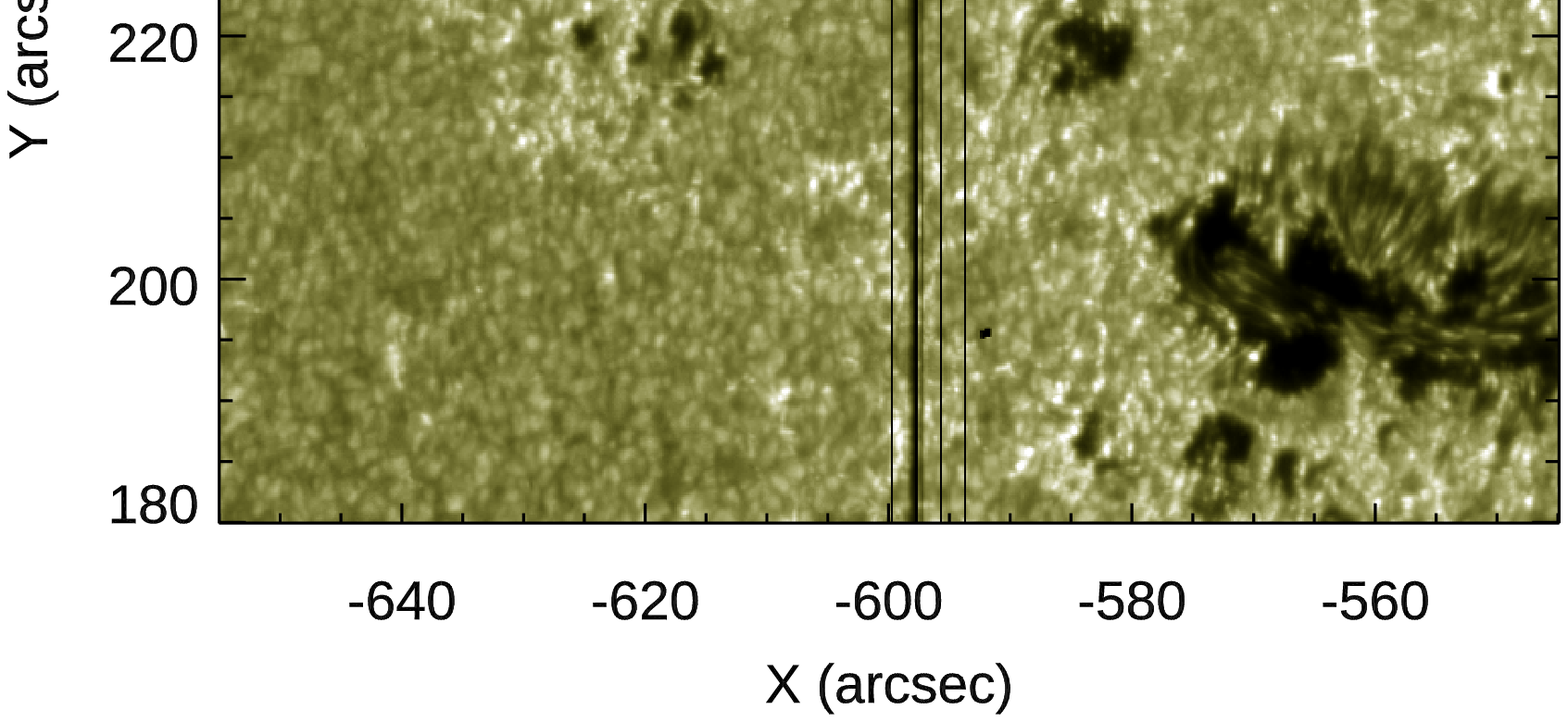}
	\includegraphics[trim=20 210 60 230, clip, scale=.39]{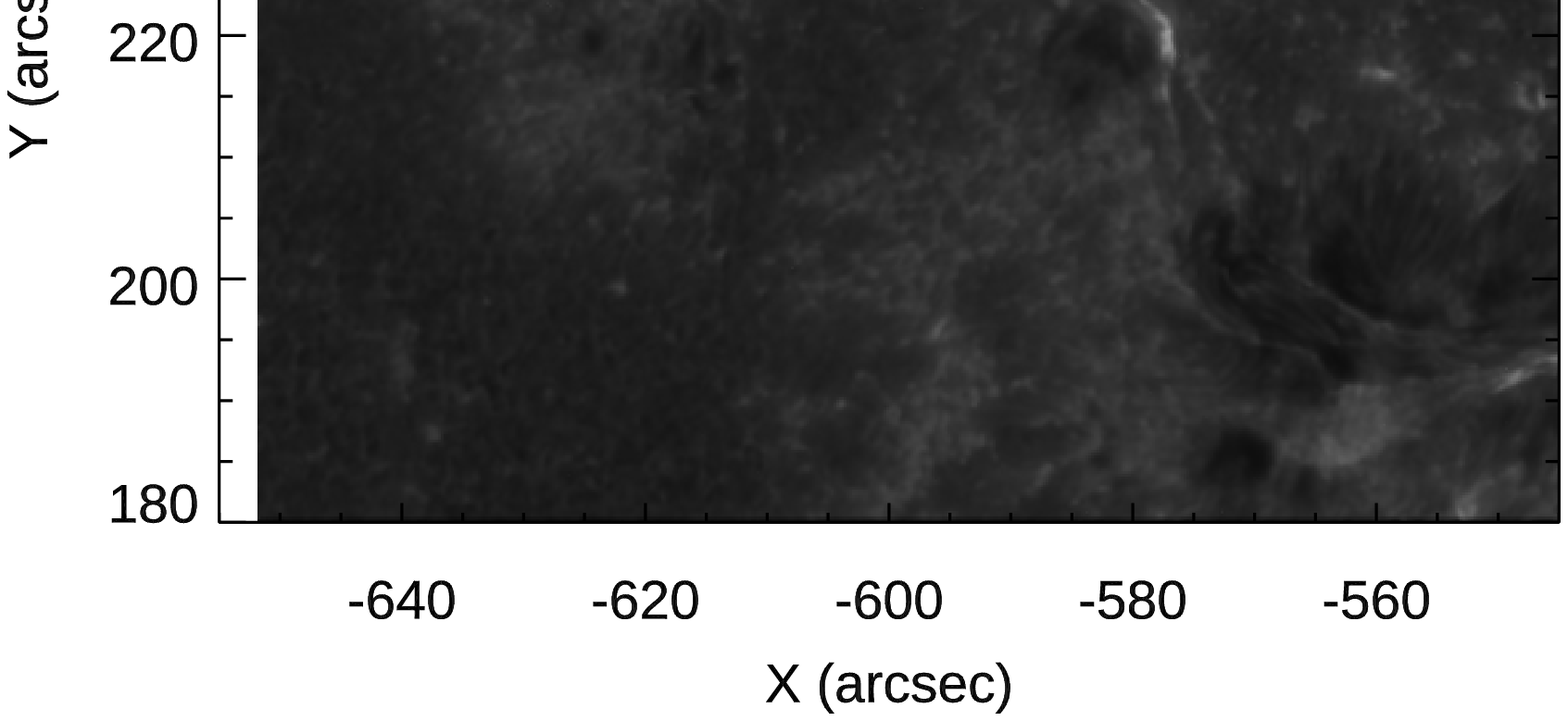}	
	\caption{\textit{Top panels}: Maps derived from \textit{IRIS} SJI images acquired on 2014 November~7 at around 16:56~UT, showing a portion of the FOV recorded by the instrument at three different wavelengths: SJI 1330~\AA{} (first panel), SJI 2796~\AA{} (second panel), and SJI 2832~\AA{} (third panel). In each image the dashed vertical lines indicate the raster positions during the acquisition time. \textit{Bottom panel}: \textit{Hinode}/SOT \ion{Ca}{2} H filtergram simultaneous to \textit{IRIS} SJI images. The solid-line box indicates the sub-FOV analyzed in Figure~\ref{fig:iris_evolution}. The red-line box frames the sub-FOV zoomed in Figure~\ref{fig:iris_zoom}. The arrow in the third panel indicates the brightening. \\
	(An animation of the \textit{Hinode}/SOT \ion{Ca}{2} H images is available in the online material.) \label{fig:iris_fov}}
\end{figure}

\begin{figure*}
	\centering
	\includegraphics[trim=0 10 260 20, clip, scale=.55]{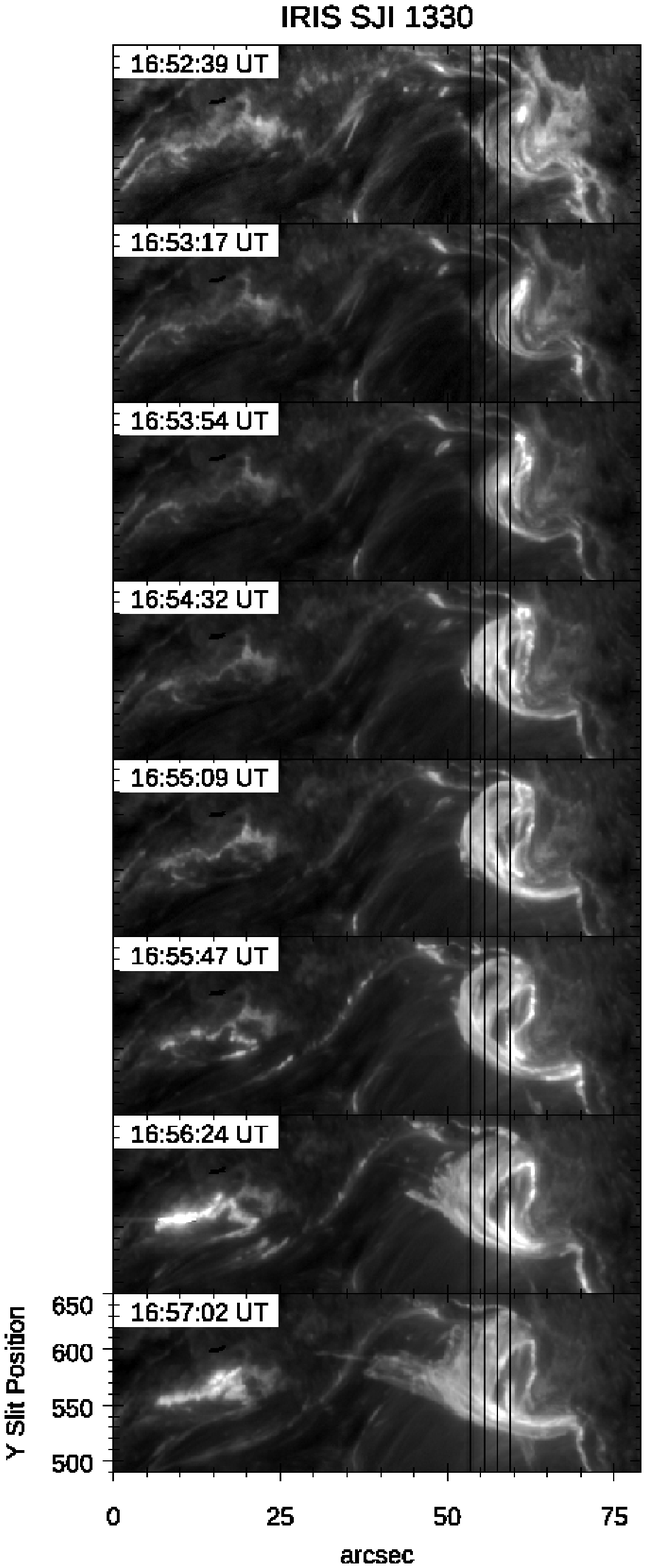}%
	\includegraphics[trim=50 10 260 20, clip, scale=.55]{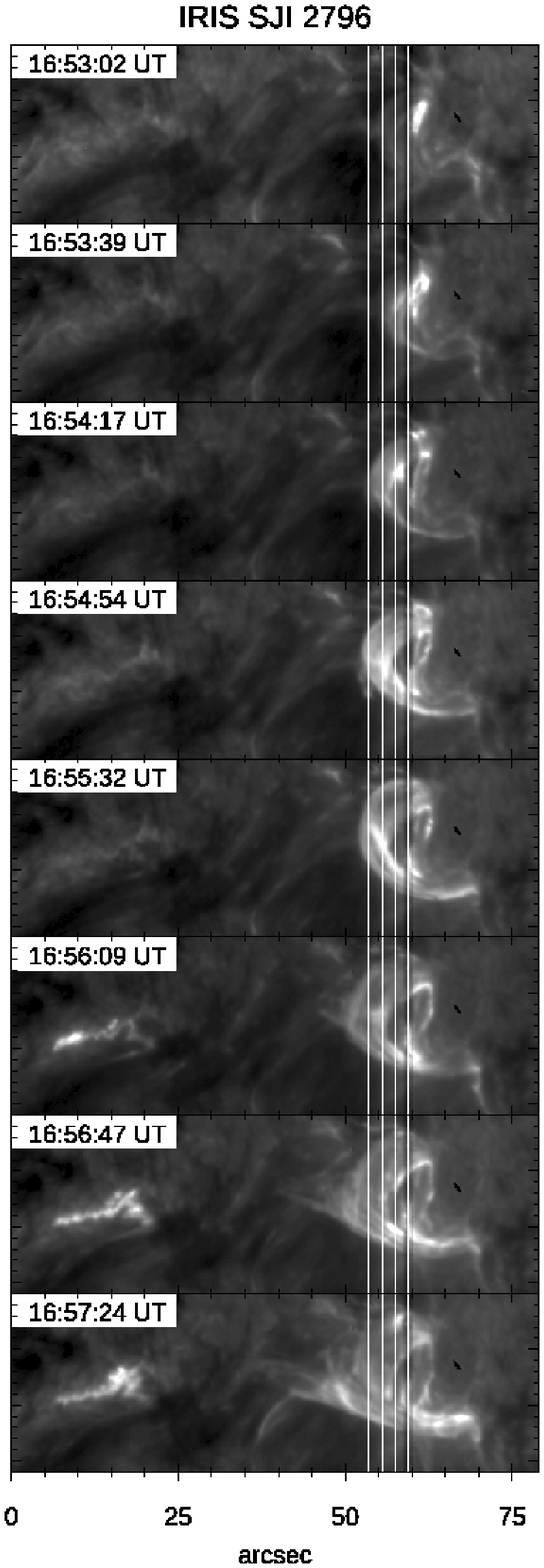}%
	\includegraphics[trim=50 10 260 20, clip, scale=.55]{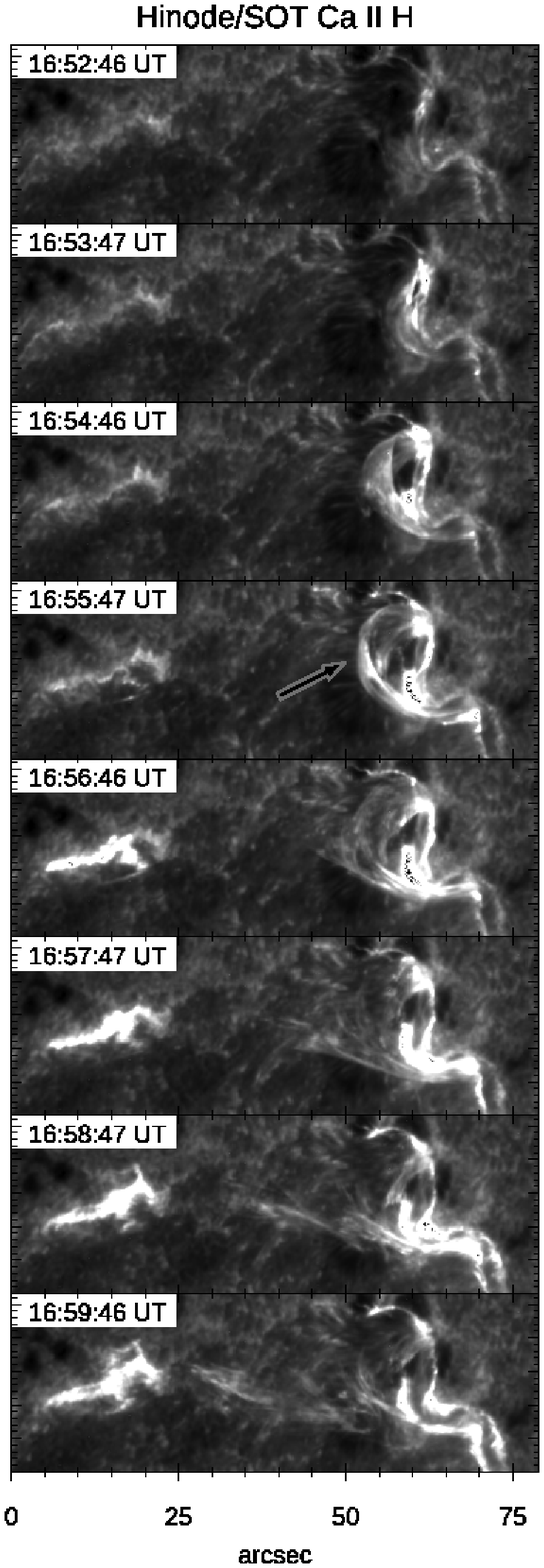}
	\caption{The evolution of the flaring site around the northern $\delta$-spot \textit{A} at the beginning of the X1.6 flare, on 2014 November~7, in three different wavelengths: \textit{IRIS} SJI 1330~\AA{} (left), \textit{IRIS} SJI 2796~\AA{} (middle), and \textit{Hinode}/SOT \ion{Ca}{2} H (right). In the \textit{IRIS} SJ images the dashed vertical lines indicate the raster positions during the acquisition time. The arrow in the \ion{Ca}{2} H filtergram indicates the expanding flux rope, which is launched during the sequence. \label{fig:iris_evolution}}
\end{figure*}

\section{Data Analysis}

As previously stated, AR~12205 was characterized by a $\beta\gamma\delta$ configuration. In particular, from the comparison of the HMI continuum image shown in Figure~\ref{fig:sdo_context} (\textit{left}) with the HMI LOS magnetogram reported in Figure~\ref{fig:sdo_context} (\textit{right}), it is possible to infer that this active region hosted two $\delta$ spots. The first one, labelled with the letter \textit{A} and indicated by the arrow pointing approximately at [-580\arcsec, 245\arcsec] in Figure~\ref{fig:sdo_context}, was characterized by the presence of penumbral filaments running almost parallel to the umbrae borders, as shown in the ROSA \textit{G}-band image reported in Figure~\ref{fig:rosa_fov} (top-right panel). The other $\delta$ spot, labelled with the letter \textit{B} and located in the region indicated by the arrow pointing at [-560\arcsec, 190\arcsec] in Figure~\ref{fig:sdo_context}, was part of an area characterized by a magnetic neutral line running almost parallel to the equator. Also in this case, the $\delta$ spot is characterized by penumbral filaments running parallel to the opposite polarity umbrae, as shown in the ROSA \textit{G}-band image reported in Figure~\ref{fig:rosa_fov} (middle panels). 

During the observing interval, two flares took place in AR~12205: a C7.0 class flare (\texttt{SOL2014-11-07T16:10}, start time 16:10~UT, peak 16:39~UT, end 16:45~UT), and an X1.6 class flare (\texttt{SOL2014-11-07T16:53}, start time 16:53~UT, peak 17:26~UT, end 18:34~UT). The dashed vertical lines in Figure~\ref{fig:goes} indicate the flare peaks. 

From an inspection of Figure~\ref{fig:sdo_context} (\textit{left}), showing the photospheric configuration of AR~12205, it is possible to infer that the SJI FOV (dashed-line square) roughly includes both ROSA FOVs, although it should be stressed that the slit position allowed us to follow only the evolution of the northern part of the flares, where $\delta$-complex \textit{A} was located.

\subsection{The C7.0 flare evolution}

Figure~\ref{fig:flareC} displays the evolution of the C7.0 flare at different atmospheric heights, from the upper photosphere (\textit{SDO}/AIA 1600~\AA{}) up to the hot corona (\textit{SDO}/AIA 131~\AA{}), including high-resolution observations by \textit{IRIS} in the upper chromosphere (SJI 2796~\AA{}) and transition region (SJI 1330~\AA{}). The FOV here analyzed covers almost the same region as in Figure~\ref{fig:sdo_context}, having been slightly extended to accommodate coronal loops of AR~12205 within the images.

At the start time of the flare (16:10~UT), a prominent dark filament is seen in the coronal channels (304, 171, 335, and 131~\AA{}) connecting the $\delta$-complex \textit{A} and the $\delta$-complex \textit{B}. While the filament is being lifted (16:15~UT), brightenings are seen from coronal levels down to the lower atmosphere (e.g., 1600~\AA{} and 2796~\AA{}) in both the $\delta$-complexes \textit{A} and \textit{B}. Some remote brightenings are also observed, for instance the one located at [-640\arcsec, 240\arcsec], which can be explained in terms of the connectivities of AR~12205 in the upper atmospheric layers (compare to the 131~\AA{} image). Energy release, likely due to small-scale reconnection episodes in the region where magnetic field lines have been pushed into by the rising filament, causes intensity enhancements in this area in the transition region (1330~\AA{} and 304~\AA{}). At 16:20~UT, there is an apparent motion of plasma toward the $\delta$-complex \textit{A} and a null point seems to form (see at [-620\arcsec, 230\arcsec] in the 171~\AA{} map). The dark filament is being shaken, but it is not launched off. The null point is still well visible at 16:25~UT ([-620\arcsec, 220\arcsec] in the 171~\AA{} and 335~\AA{} maps), while at the base of the dome-shaped domain under the null point a bright patch at X=(-580\arcsec), elongated along the Y direction, is seen in all the channels, down to the chromosphere. In the full sequence, this bright patch is seen moving southward. At the peak, the C7.0 flare exhibits a circular ribbon, which occurs in the area of the $\delta$-complex \textit{B}. The dark filament has settled down, acquiring a sigmoidal configuration along the PIL of the AR (see the 304~\AA{} maps at 16:39~UT and 16:45~UT). Shortly after the C7.0 flare ends, the X1.6 flare starts (16:53~UT).

\begin{figure}
	\centering
	\includegraphics[trim=20 210 80 200, clip, scale=.4]{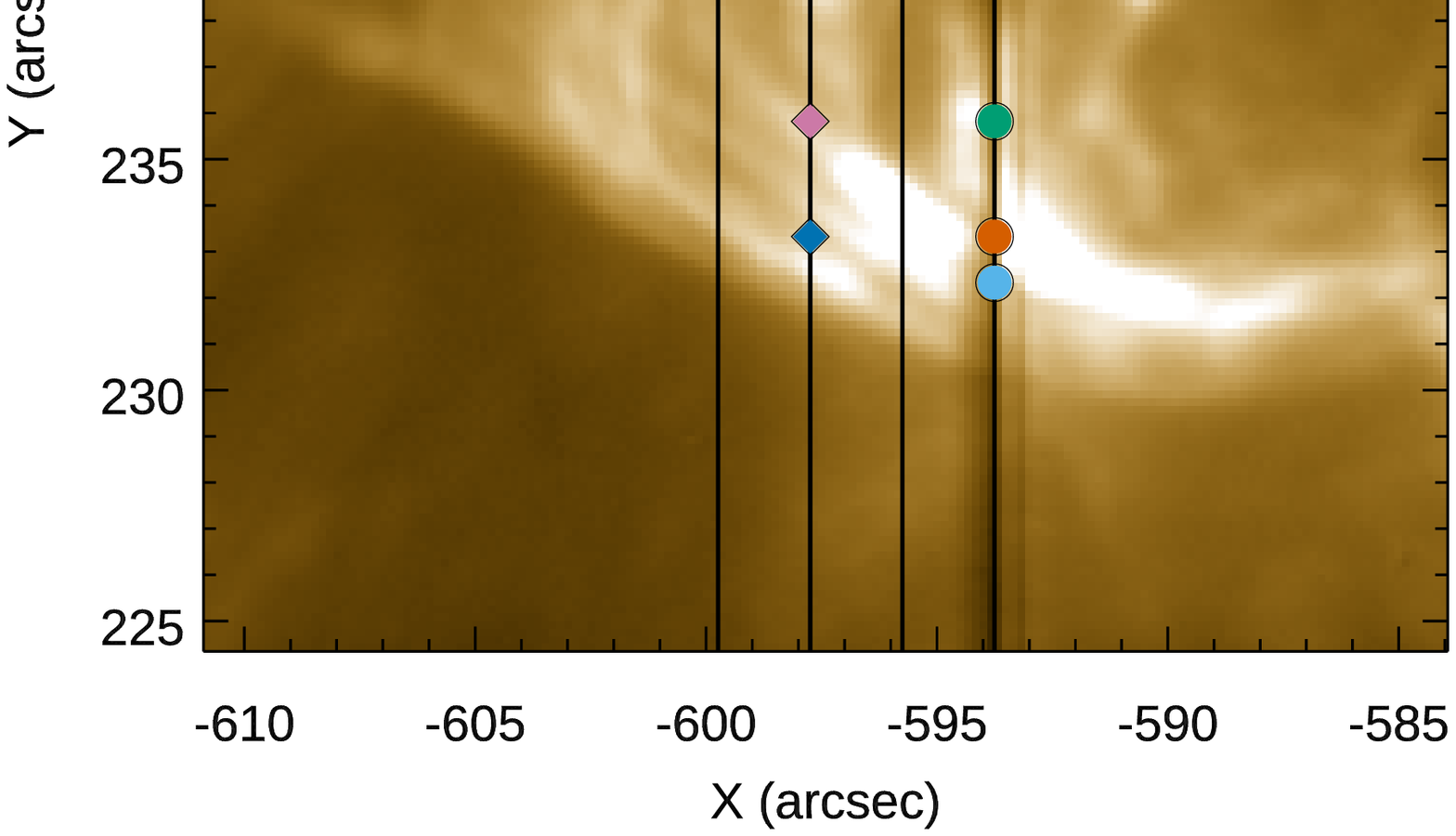}  
	\includegraphics[trim=20 150 80 200, clip, scale=.4]{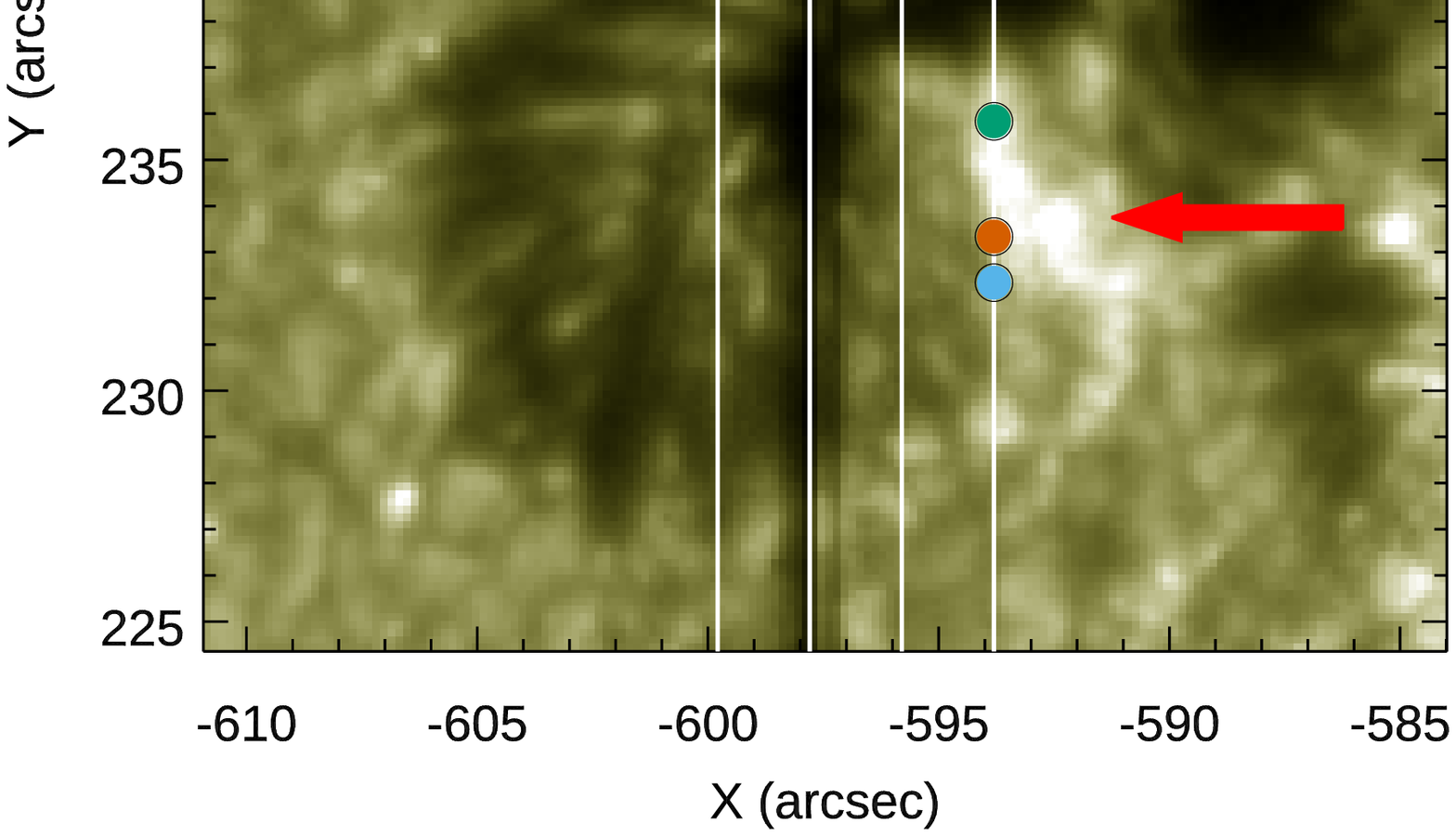}
	\caption{Zoomed maps at around 16:56~UT showing the details of $\delta$-spot \textit{A}, characterized by the presence of sheared penumbral filaments within the two opposite magnetic polarities. The sub-FOV is indicated with a dashed-line box in Figure~\ref{fig:iris_fov}. The blue, orange, and green circles indicate the slit positions used to determine the line profiles shown in Figures~\ref{fig:iris_continua} and~\ref{fig:iris_profiles}. Analogously, the colored diamonds indicate the slit positions of the profiles shown in Figure~\ref{fig:iris_fluxrope}. The arrow indicates the brightening. \label{fig:iris_zoom}}
\end{figure}

\subsection{The X1.6 flare evolution}

In the following, we will describe the evolution of the main X1.6 flare and the several phenomena that can be recognized during its occurrence.

\subsubsection{The onset of the flare}

Figure~\ref{fig:iris_fov} displays the maps obtained from the \textit{IRIS} SJIs acquired in the three wavelengths: SJI 1330~\AA{} (first panel), SJI 2796~\AA{} (second panel), and SJI 2832~\AA{} (third panel). The portion of the simultaneous \textit{Hinode}/SOT \ion{Ca}{2} H filtergram cospatial to the \textit{IRIS} maps is also displayed (fourth panel). We indicate with boxes the subfields of the full FOV (sub-FOVs) used for the subsequent analysis.

In Figure~\ref{fig:iris_evolution} we plot a time sequence, of the sub-FOV indicated with a solid-line box in Figure~\ref{fig:iris_fov} containing the northern $\delta$-complex \textit{A}, composed by SJIs at 1330~\AA{} (first column) and 2796~\AA{} (second column), and the nearly simultaneous cospatial \textit{Hinode}/SOT \ion{Ca}{2} H images (third column). The sequence, referring to the start time of the X1.6 flare (16:53~UT) and following instants, shows that the flare initiated between the two opposite-polarity umbrae of the $\delta$-sunspot \textit{A}. Both \textit{IRIS} and \textit{Hinode}/SOT observed the emergence and subsequent expansion of bright loops from the $\delta$-sunspot. These formed an expanding flux rope, as indicated by the arrow in the \ion{Ca}{2} H filtergram at 16:55~UT. Notably, at around 16:56~UT, this loop systems broke up into two different branches. \citet{yurchy15} observed the development of these structures also in \textit{SDO}/AIA images. Also, we notice prior to the breaking of the bright loops a third luminous elongated patch appears about 30\arcsec{} to the East of the loop systems, visible in \textit{IRIS} 1330~\AA{}, 2796~\AA{}, and even more evident in the \ion{Ca}{2} H filtegrams. While \textit{IRIS} observations abruptly end just a few moments after the breaking of the loop system and flux rope ejection, \ion{Ca}{2} H images follow the subsequent evolution of the flare: the eastern elongated patch becomes stronger in intensity, growing in size and slightly moving. At 16:59~UT, the loop system is definitely broken apart and the two ribbons clearly appear. The evolution of the onset of the flare can be also seen in the online movie for the \textit{Hinode}/SOT \ion{Ca}{2} H images relevant to Figure~\ref{fig:iris_fov}.

\subsubsection{Enhancements in the continuum and in lines}
At around 16:56~UT, that is, at the beginning of the rise phase of the X1.6 flare, the \textit{IRIS} SJI image in the \ion{Mg}{2} wing at 2832~\AA{} shows an intensity enhancement in the site at X=(-595\arcsec, -590\arcsec), Y=(230\arcsec, 235\arcsec) in Figure~\ref{fig:iris_fov} (\textit{third panel}). This intensity enhancement has the shape of a small arch that moves toward the south-east direction, encountering therefore the place monitored by the \textit{IRIS} slit (see also Figure~\ref{fig:iris_zoom}, bottom panel). Unfortunately, the \textit{IRIS} satellite after this time changed its target, so we do not have a complete coverage of the phenomena occurring during this flare. Nevertheless, it has been possible to study the intensity enhancement and the plasma motions at the time when the slit was on the flare ribbon. 

At around 17:22~UT, i.e., a few minutes before the peak of the X1.6 flare, WL ribbons were detected close to the southern $\delta$ spot \textit{B} in both the \textit{G} band and the 4170~\AA{} continuum images acquired with ROSA (FOV2) (see also the online movie for Figure~\ref{fig:rosa_fov}). In particular, the southernmost ribbon is clearly visible in the images obtained in these wavelength ranges, while the northern ribbon can be distinguished only using the difference imaging (see, e.g., Figure~\ref{fig:rosa_fov}, bottom-left panel). The ribbons separate with a velocity of $\approx 10 \,\mathrm{km\,s}^{-1}$. In Figure~\ref{fig:rosa_fov} (bottom-right panel) we plot the lightcurve obtained using the \textit{G}-band sequence. It has been calculated at the location of the flare ribbons, in the upper half of zoomed FOV2 where the flare ribbons are located. It clearly illustrates that an intensity enhancement is found almost simultaneously with the flare peak.

\begin{figure*}[t]
	\centering
	\includegraphics[trim=20 20 0 10, clip, scale=0.6]{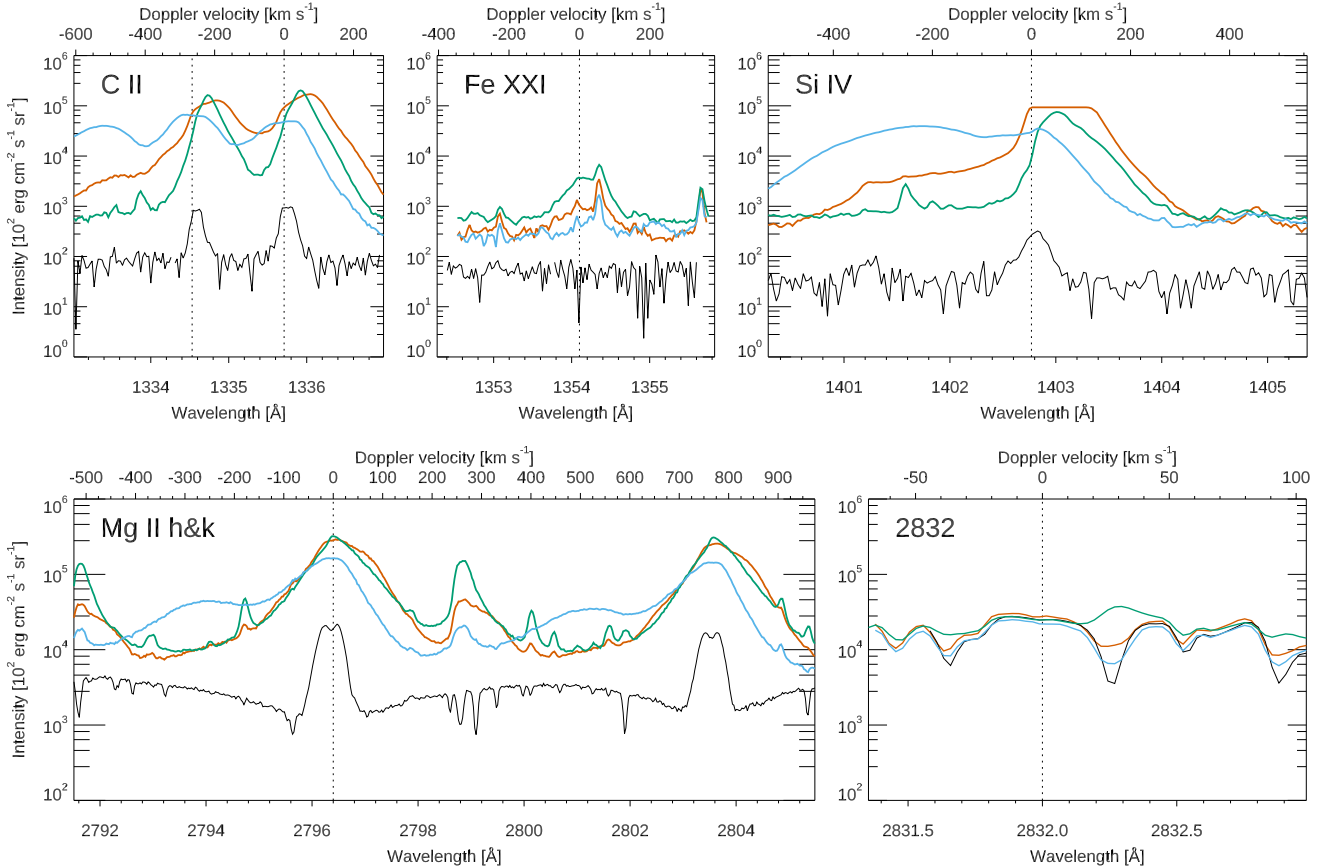}
	\caption{Average intensity as a function of wavelength in five of the IRIS spectral windows in the pixels at raster positions (3,[537:539]; blue), (3,[543:545]; orange), and (3,[558:560]; green) at 16:56:47~UT, corresponding to (-593\farcs8, 232\farcs3), (-593\farcs8, 233\farcs3), and (-593\farcs8, 235\farcs8), respectively. \textit{Black line}: The average intensity calculated at the same time along the 20 consecutive slit positions (from 160 to 179), corresponding to a quiet-Sun region.
	\label{fig:iris_continua}}
\end{figure*}

\begin{figure*}
	\centering
	\includegraphics[trim=0 10 0 60, clip, scale=0.6]{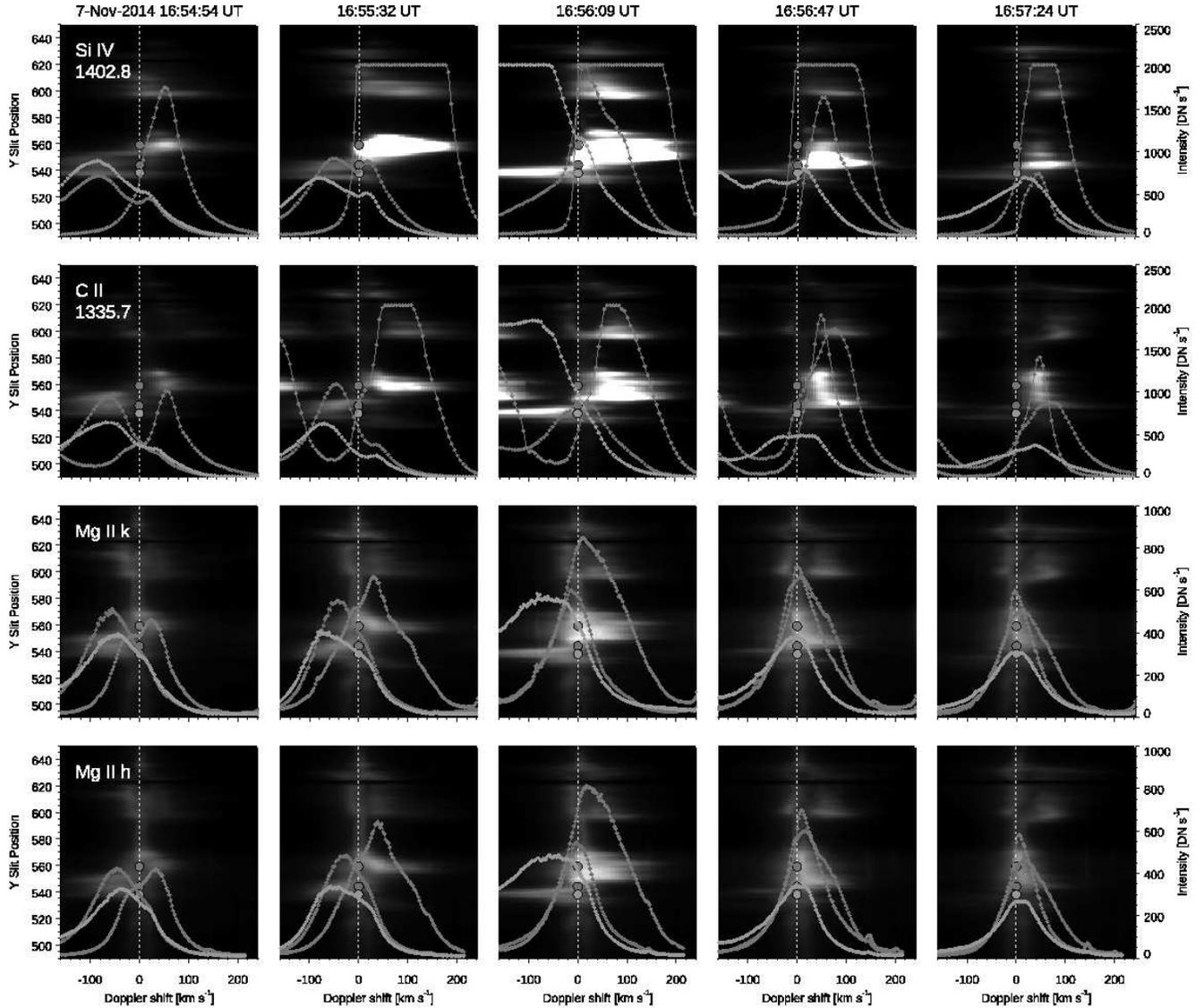}   
	\caption{Line profiles for three different pixel positions of the IRIS slit for \ion{Si}{4} 1402.8~\AA{} (top row), \ion{C}{2} 1335.75~\AA{} (second row), \ion{Mg}{2}~k 2796.31~\AA{} (third row) and \ion{Mg}{2}~h 2803.55~\AA{} (bottom row). In each row the line profiles for successive times are overplotted on the relevant spectrograms. The dashed vertical lines indicate the position of the line center, while the blue, orange, and green circles show the slit positions relevant to the profiles indicated with the same colors. 
	\label{fig:iris_profiles}}
\end{figure*}

\begin{figure*}
	\centering
	\includegraphics[trim=0 10 0 60, clip, scale=0.6]{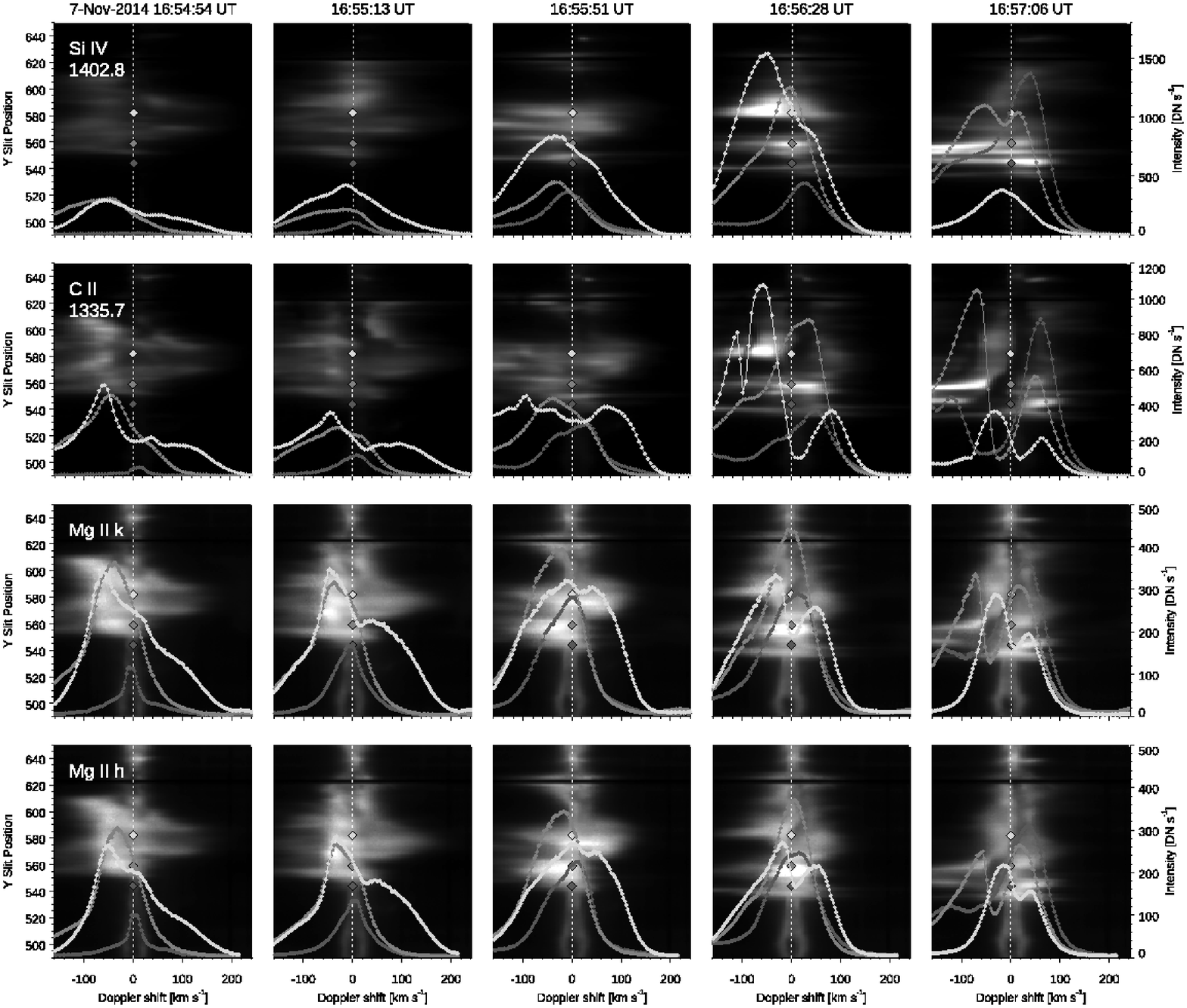}   
	\caption{Same as in Figure~\ref{fig:iris_profiles}, for the slit positions indicated with the blue, magenta, and yellow diamonds in Figure~\ref{fig:iris_zoom}. 
	\label{fig:iris_fluxrope}}
\end{figure*}

In Figure~\ref{fig:iris_continua} we display the radiometric calibrated intensities in five \textit{IRIS} spectral windows expressed in physical units ($\mathrm{erg}\:\mathrm{cm}^{-1}\:\mathrm{s}^{-1}\:\mathrm{sr}^{-1}$ per pixel), which have been obtained from the sub-FOV with the \textsc{iris\textunderscore{}calib\textunderscore{}spec} routine. Profiles shown in blue refer to the average intensity in the pixel at slit position (3,538) with its adjacent pixels (3,537) and (3,539), relevant to raster 79. The approximate position of these pixels is shown in Figure~\ref{fig:iris_zoom}, with a blue circle; the other circles (orange and green) shown in the same Figure~\ref{fig:iris_zoom} refer to the other pixels studied in this work. The position indicated by the blue circle corresponds to [-593\farcs8, 232\farcs3] at 16:56:47~UT, at the beginning of the X1.6 flare. Analogously, profiles shown in orange refer to the average intensity in the pixel at slit position (3,544) with its adjacent pixels (3,543) and (3,545) for the same raster. This position corresponds to [-593\farcs8, 233\farcs3]. Finally, profiles shown in green refer to the average intensity in the pixel at slit position (3,559), with its adjacent pixels (3,558) and (3,560) for the same raster, the position corresponding to [-593\farcs8, 235\farcs8] solar coordinates. For comparison, in the same plot the average intensity calculated at the same time along the slit at 20 consecutive pixel positions (from 160 to 179), in a quiet-Sun region, is shown with a black solid line. 

It is evident that in all the channels there is an intensity enhancement also in the continuum region, with the exception being the 2832~\AA{} spectral window. Nevertheless, also in the latter we see emission in the flaring pixel at (3,559) corresponding to absorption features in the quiet Sun. 

Interestingly, the blue pixel exhibits a very prominent bump in the blue wing of the \ion{Si}{4} 1402~\AA{} line, as well as in the blue wings of the \ion{C}{2} 1334 and 1336, and \ion{Mg}{2}~h and~k lines. It is worth highlighting that this slit position corresponds to the apparent point where the loop systems studied in Figure~\ref{fig:iris_evolution} splits into two branches, as clearly seen in Figure~\ref{fig:iris_zoom} (top panel).

In Figure~\ref{fig:iris_profiles} we show a sample of the line profiles at different pixel positions of the \textit{IRIS} slit, where the bright ribbon close to $\delta$-spot \textit{A} was observed. 

The analysis of the \ion{Si}{4} 1402.8~\AA{} profiles reported in Figure~\ref{fig:iris_profiles} (top panels) indicates that at the position of the green pixel, during the analyzed time interval there is a strong downflow with velocities reaching $100$~\kms{}; the line is saturated at 16:55:32 and 16:56:09~UT. The profiles relevant to the orange pixel show a sudden change from upflows ($-90$~\kms) to downflows (up to $80$~\kms); the line shows a double peak at 16:55:32~UT and becomes saturated at 16:56:47 and 16:57:24~UT. The blue pixel has profiles whose behaviour is quite similar to that of the orange profile. However, in the former there is a sudden increase of the upflows (up to $-100$~\kms) at 16:56:09~UT. At 16:57:24~UT downflows of $\approx 25$~\kms{} are detected.

If we examine the profiles reported in Figure~\ref{fig:iris_profiles} (second panels), relevant to the \ion{C}{2} 1335.75~\AA{} line, we can infer a similar behaviour to that we found in the \ion{Si}{4} 1402.8~\AA{} line. In fact, the profiles relevant to the green pixel exhibit downflows with velocities ranging between $60$ and $100$~\kms; moreover, the line is saturated at 16:55:32 and 16:56:09~UT. The orange pixel shows a sudden change from upflows ($-50$~\kms) to downflows (up to $100$~\kms), similarly to the \ion{Si}{4} 1402.8~\AA{} line. In this case, the line is saturated at 16:56:09~UT. The blue profile is akin to the orange profile, except for the sudden increase of the upflow, up to $-100$~\kms, observed at 16:56:09~UT.

Finally, the profiles of the \ion{Mg}{2}~h\&k lines (shown in the bottom panels of Figure~\ref{fig:iris_profiles}) indicate that during the analyzed time interval at the green position there is evidence of downflows with velocities ranging between $35$ and $40$~\kms{} at the beginning; later on, these motions are not evident anymore. The orange profiles are initially characterized
by an upflow ($-65$~\kms), which later disappears. The behaviour of the blue profiles is rather similar to that found in the orange profiles, but at this position a sudden increase of the upflows (up to $-100$~\kms) occurs at 16:56:09 UT.

In Figure~\ref{fig:iris_fluxrope} we analyze the flux rope ejected at the beginning of the X1.6 flare. The slit sequentially crosses expanding sections of the ejected flux rope. This is clearly shown as the emitting area across the slit in the spectrograms widens with time.

Using a similar approach as in Figure~\ref{fig:iris_profiles}, we investigate the behaviour of the UV emission in some pixel positions relevant to the expanding flux rope in Figure~\ref{fig:iris_fluxrope}. We indicate the approximate position of these pixels in Figure~\ref{fig:iris_zoom} (top panel) with colored diamonds. In particular, the slit position (1,544) indicated by the blue diamond corresponds to [-597\farcs8, 233\farcs3] solar coordinates at 16:56:47~UT, the position (1,560) to [-597\farcs8, 235\farcs8] (magenta diamond), and the position (1,582) to [-597\farcs8, 239\farcs6] (yellow diamond). We see that the yellow and magenta pixels exhibit upflows of about $-50$~\kms since 16:54:54~UT. At the beginning, lines at chromospheric heights (\ion{C}{2} and \ion{Mg}{2}~h\&k) are stronger with respect to the background than the \ion{Si}{4} line is. At the position indicated by the blue diamond, we see the intensity increase of the line profiles, indicating that the flux rope reaches that slit position and hotter plasma is observed. At 16:55:51~UT, we see a blue asymmetry in chromospheric lines. Interestingly, at 16:56:28~UT we observe strong blue bumps with a blueshift of $\approx -60$~\kms{} in the \ion{Si}{4} and \ion{C}{2} lines. In the latter, we notice a blend with an absorption line, located at $\Delta \lambda \approx -80$~\kms, as well as a gap near the line center that moves from a blueshifted position at $\mathrm{Y}\approx 580$  to a redshifted position at increasing $\mathrm{Y}$ values, until $\mathrm{Y}\approx 620$. A similar behaviour is observed at 16:57:06~UT, with a change from $\Delta \lambda \approx -40$~\kms{} to $\Delta \lambda \approx +20$~\kms. At this time, we see different plasma components in the \ion{Si}{4} and \ion{C}{2} lines, with one component characterized by upflows up to $-80$~\kms{} and the other by moderate downflows of a few tens of \kms. However, in all the lines the blue pixel exhibits downward motions of about $+20$~\kms{} at 16:57:06~UT. In the bulk of the chromosphere (\ion{Mg}{2}~h\&k), at both 16:56:28~UT and 16:57:06~UT we find some locations with blueshifted mustaches in the spectroheliograms. 

Unfortunately, the \textit{IRIS} observations stop at this point, thus we cannot longer follow the expansion of the flux rope.

\subsection{Changes in the penumbrae}

\begin{figure}[t]
	\centering
	\includegraphics[trim=50 10 20 120, clip, scale=0.285]{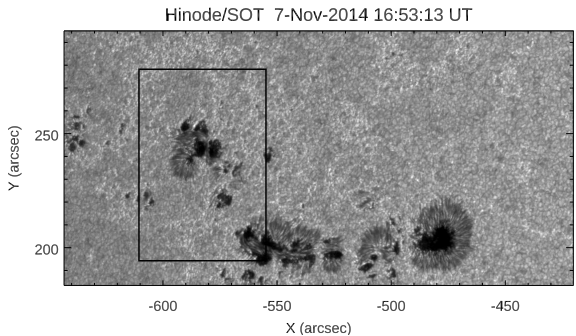}
	\caption{Map of the \textit{G}-band intensity taken by \textit{Hinode}/SOT at the beginning of the X1.6 flare. The box frames the sub-FOV used in Figure~\ref{fig:sot}, relevant to the $\delta$ complex $A$. \label{fig:sot_context}}
\end{figure}

\begin{figure*}[t]
	\centering
	\includegraphics[trim=40 70 120 10, clip, scale=0.65]{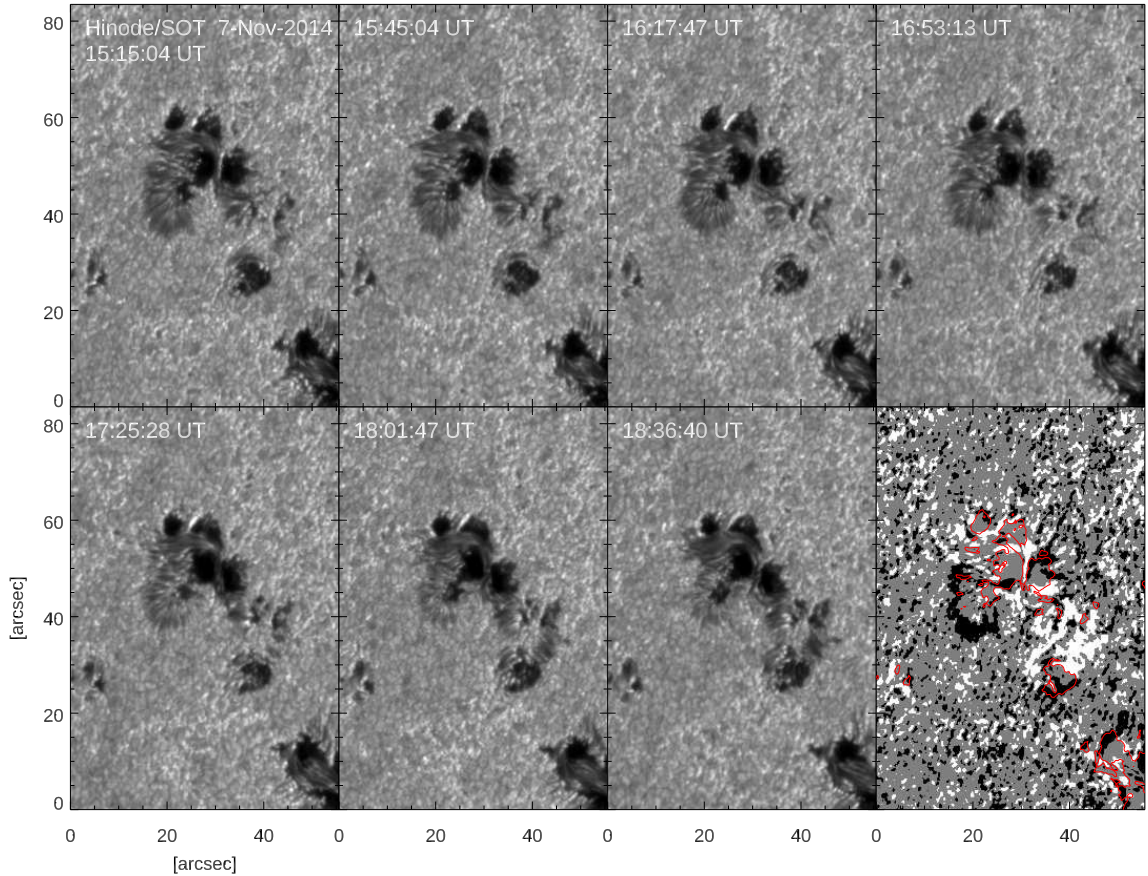}
	\caption{\textit{Panels 1--7}: Photospheric evolution of the $\delta$ complex $A$ during the C7.0 and X1.6 flares, as seen by \textit{Hinode}/SOT in \textit{G} band. \textit{Bottom-right panel}: Difference image between the first and last \textit{G}-band filtergrams of the \textit{Hinode}/SOT sequence. White (black) areas indicate regions with penumbral enhancement (decay). Red contours indicate the umbral boundary at the beginning of the sequence. \label{fig:sot}}
\end{figure*}

Figure~\ref{fig:sot_context} shows AR~12205 at the beginning of the X1.6 flare, as observed by \textit{Hinode}/SOT in the \textit{G} band. We indicate with a solid box the sub-FOV relevant to the flaring $\delta$ complex $A$ observed to the north-east of AR~12205. The sequence of \textit{Hinode}/SOT filtergrams displayed in Figure~\ref{fig:sot} manifests the concurrence of stable penumbral decay and enhancement in this area, during the evolution of the observed flares. In particular, after the X1.6 flare, the penumbra surrounding the eastern part of negative spot of the $\delta$ complex $A$ is intensely reduced, whereas penumbral filaments are enhanced along the polarity inversion line to the north of the same $\delta$ complex and, above all, to the north of the proto-spot in the south-western part of the sub-FOV of these filtergrams. The difference image between the first and last filtergrams in the sequence (Figure~\ref{fig:sot}, bottom-right panel) clearly shows the areas of permanent penumbral decay (black) and enhancement (white). Note that we apply a mask, so that the black/white areas refer to regions where the normalized continuum intensities changed more than $\pm 0.15$, in absolute value.

\begin{figure*}[t]
	\centering
	\includegraphics[trim=10 0 20 260, clip, scale=0.625]{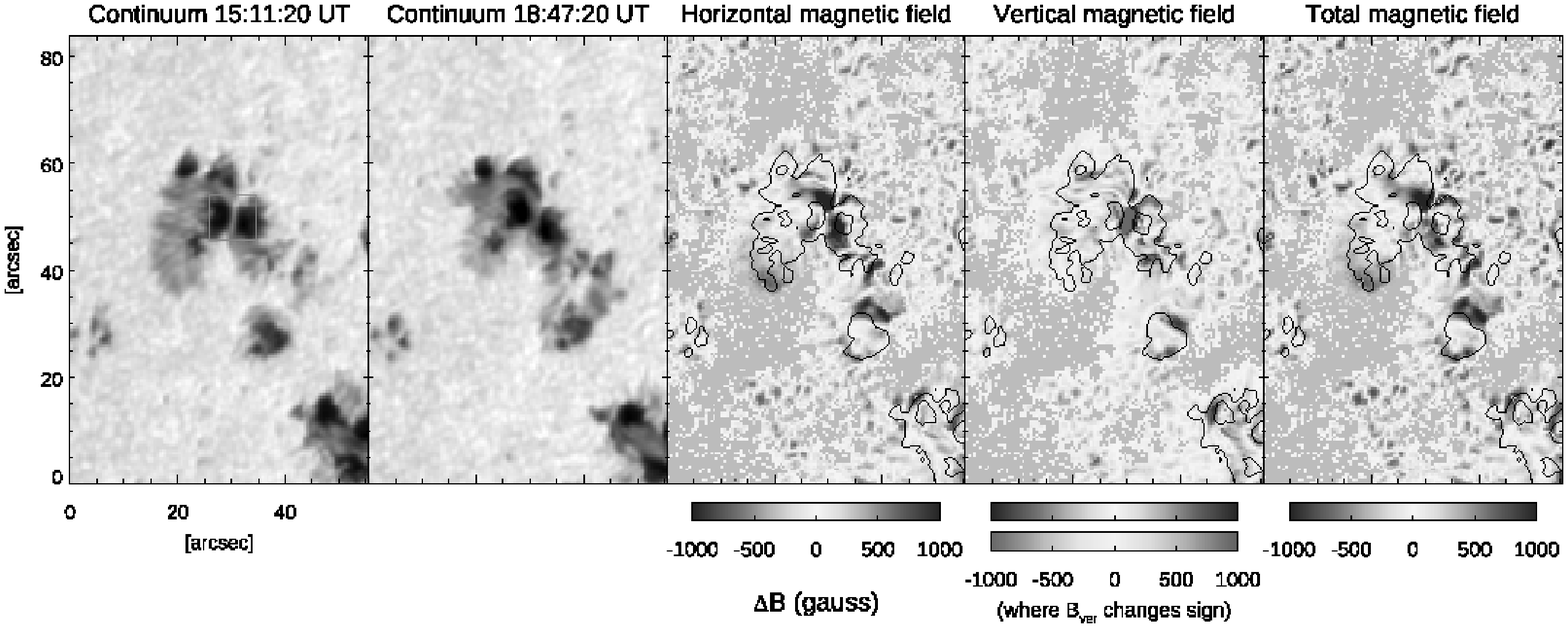}
	\caption{\textit{First and second panels:} Continuum images relevant to the same sub-FOV as used for the analysis of \textit{Hinode}/SOT filtergrams (see Figure~\ref{fig:sot}), for \textit{SDO}/HMI before (15:11~UT) and after (18:47~UT) the flares. \textit{Third, fourth, and fifth panels:} Difference images between the final and initial maps of horizontal field component, vertical field component, and total magnetic field strength. The grey background indicates pixels with total magnetic field strength $<100$~G, not considered. The color bars represent the variation of the magnetic field components and of the total field strength (blue: $-1000$~G; red: $1000$~G). The extra color bar indicates the variation of the vertical component of the magnetic field where it changes sign (yellow: $-1000$~G; green: $1000$~G). \label{fig:magnetic}}
\end{figure*}

Using \textit{SDO}/HMI SHARPs data with 12 minutes cadence \citep{Bobra:14}, we were also able to obtain information about the magnetic field configuration in the region around $\delta$ complex $A$, where penumbral decay/enhancements occur.

Figure~\ref{fig:magnetic} shows the same sub-FOV as Figure~\ref{fig:sot}. The first two panels display the continuum maps from \textit{SDO}/HMI before (15:11~UT) and after (18:47~UT) the C7.0 and X1.6 flares, where one can easily recognize the areas where penumbrae have disappeared or have newly formed. The remaining panels of Figure~\ref{fig:magnetic} are the difference images for the horizontal field component (third panel), vertical field component (fourth panel), and total magnetic field strength (fifth panel). The difference images were calculated sub-FOV the magnetic field maps simultaneous with the continuum maps at 15:11~UT and 18:47~UT, respectively, as initial and final images. In these panels, we can see that major variations occur in the horizontal field component, which increases in the areas with penumbral enhancement ($\Delta B$ up to $1000$~G) and, conversely, decreases in areas where penumbrae decay. This behaviour is reflected in the variations of the total magnetic field strength as well. The changes in the vertical field component are smoother, except for a small area around the polarity inversion line of the $\delta$ complex, where a strong variation larger than $1000$~G occurs, also leading to a change in the magnetic polarity of the area.

\begin{figure}[!b] 
	\centering
	\includegraphics[trim=75 75 12 180, clip,scale=.315]{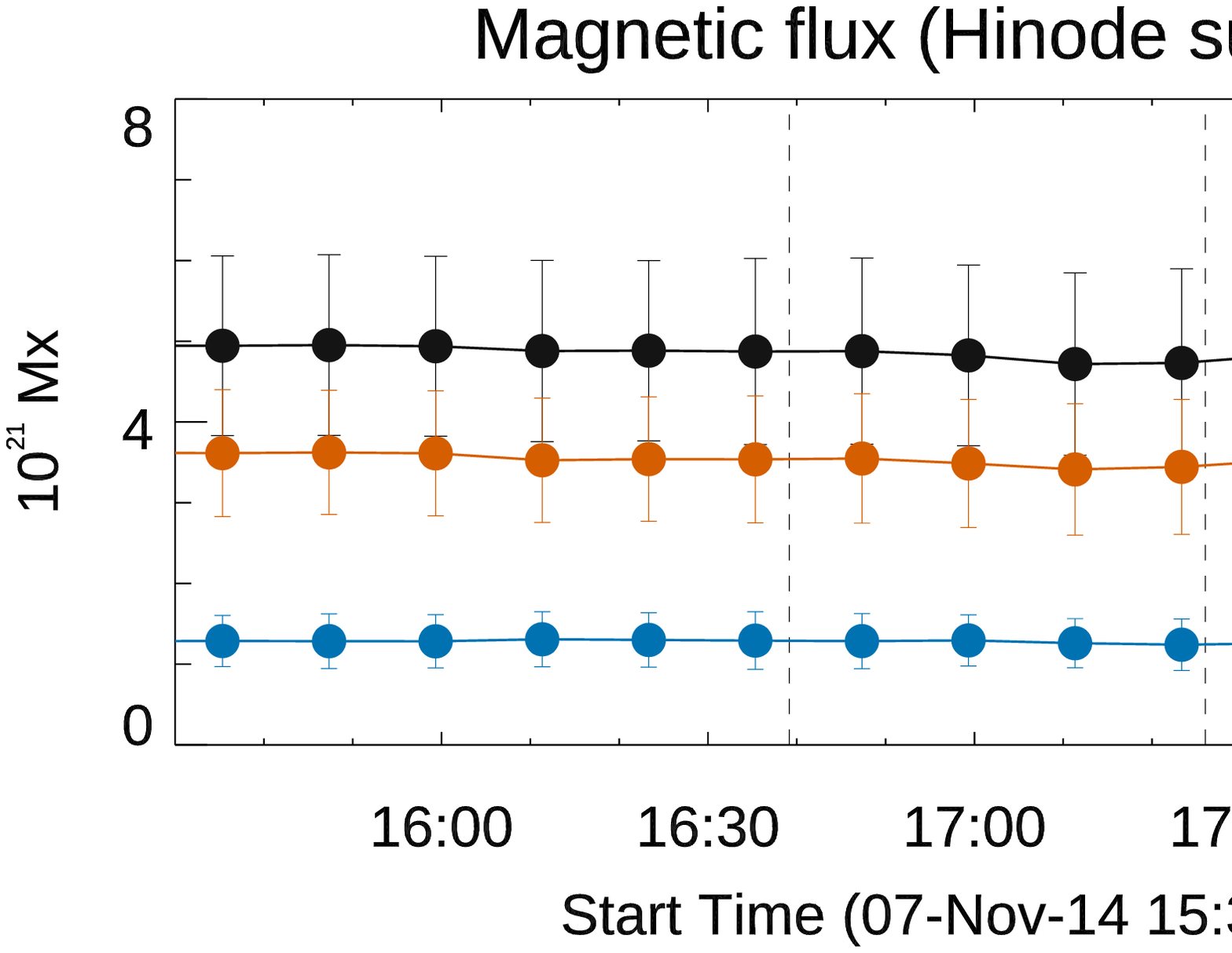} 
	\includegraphics[trim=75 75 12 180, clip,scale=.315]{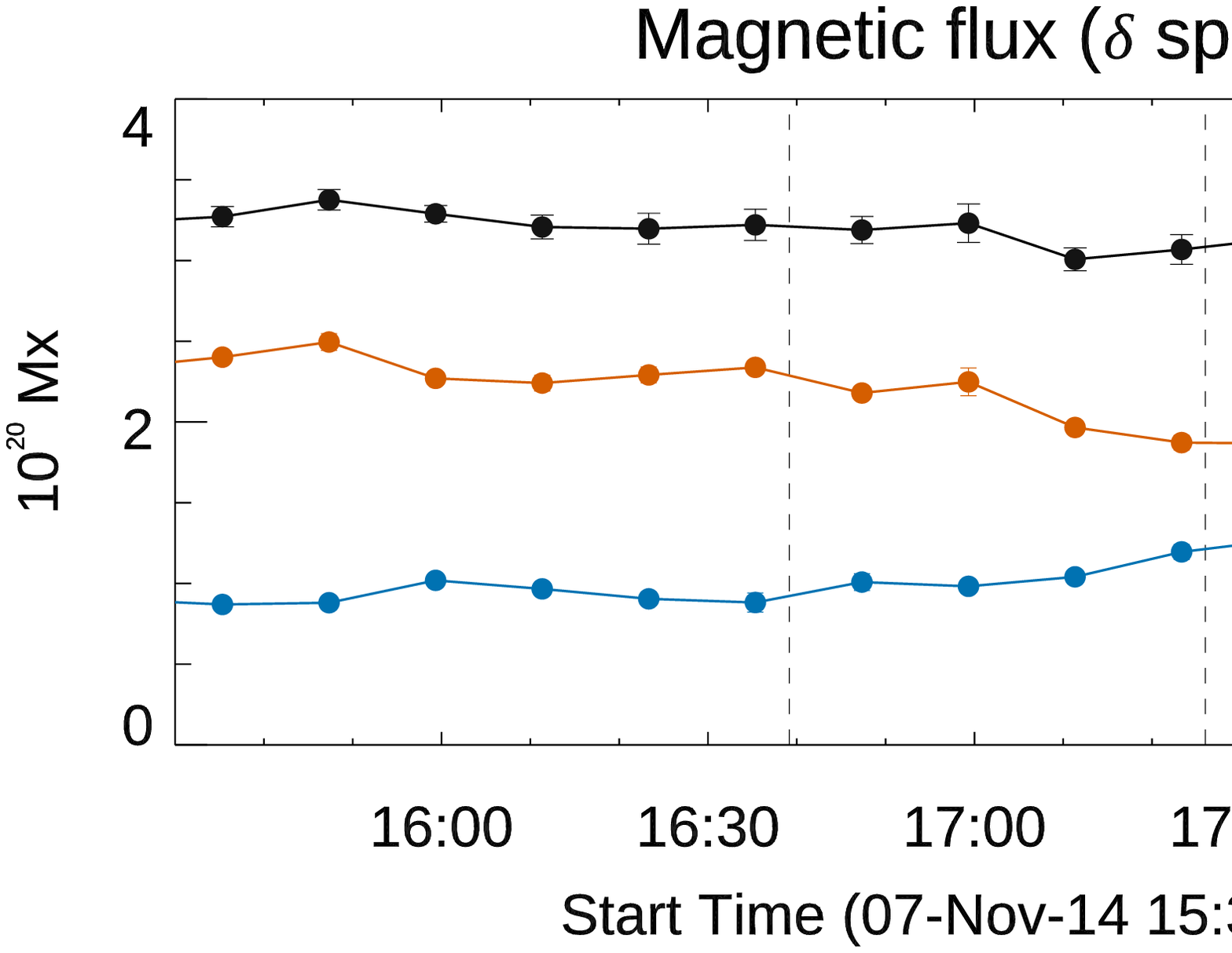} 
	\includegraphics[trim=75 45 12 180, clip,scale=.315]{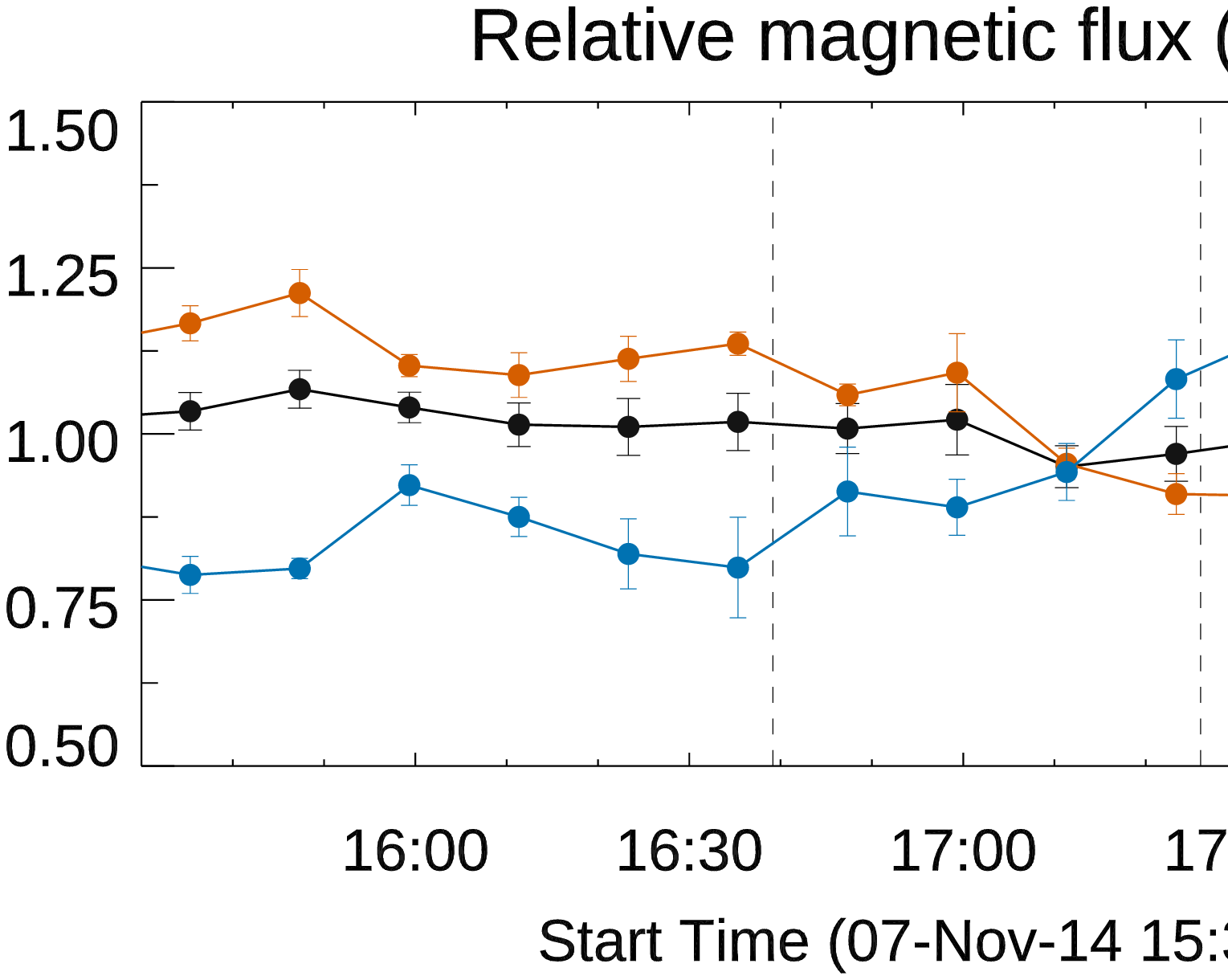} 
	\caption{Top panel: The total unsigned magnetic flux (black), positive flux (orange), and negative flux (blue, in absolute value), relevant to the \textit{Hinode} sub-FOV around $\delta$-spot \textit{A} shown in Figure~\ref{fig:magnetic}. Middle panel: Same, for the sub-FOV relevant to the inner part of the $\delta$ spot, indicated with a box in the first panel of Figure~\ref{fig:magnetic}. Bottom panel: The relative change of the total unsigned magnetic flux (black), positive flux (orange), and negative flux (blue, in absolute value), with respect to the average values, the sub-FOV relevant to the inner part of the $\delta$ spot. The peak time of the C7.0 (16:39~UT) and X1.6 (17:26~UT) flares are indicated by vertical dashed lines. Error bars represent 1-$\sigma$ value. \label{fig:flux}}
\end{figure}

The graph displayed in Figure~\ref{fig:flux} (top panel) illustrates that all the variations observed in the FOV corresponding to the difference images for the magnetic field component occur while the magnetic flux  is almost constant. Only a slight decrease takes place between the two flares. On the other hand, if we focus our attention on a smaller FOV, indicated by an orange rectangle in Figure~\ref{fig:magnetic} (first panel), relative to the region occupied by the $\delta$-complex, we notice that the positive magnetic flux (magenta circles) decreases during the selected time interval, while the negative flux (blue circles) shows an increase (Figure~\ref{fig:flux}, middle panel). This behavior is even more evident if we consider the relative magnetic flux (see Figure~\ref{fig:flux}, bottom panel). Nevertheless, the total magnetic flux remains almost constant, with a modest decrease after around 17:00. At that time, which corresponds to the onset of the X1.6 flare, the increase of negative flux and the decrease of the positive flux begin.

\section{Discussion}
In this Section we discuss the main results obtained from the study of the consecutive C7.0 [\texttt{SOL2014-11-07T16:10}] and X1.6 [\texttt{SOL2014-11-07T16:53}] flares that occurred in AR~12205 on 2014 November 7, focusing on the following aspects: 1) flare triggering; 2) emission in the continuum and line profiles; and 3) changes in the penumbrae. We shall also demonstrate how these three aspects can be linked.

\subsection{Flare triggering}

Small-scale energy releases, known as precursors, are often observed as pre-flare brightenings before the onset of large flares \citep{Wang:17}. In the case of AR~12205, a serie of minor flares occurred on 2014 November~7 before the X1.6 flare [\texttt{SOL2014-11-07T16:53}]. Indeed, \citet{Sobotka:16} observed at high resolution the occurrence of some brightenings during the C3.9 flare with peak at 12:03~UT in the $\delta$-complex \textit{B}, which hosted the circular ribbon of the flare. In addition, they saw signatures of slipping reconnection appearing as flare-like brightenings that first occurred in the $\delta$-complex \textit{A} and then occurred southward and reached the $\delta$-complex \textit{B}, where the C3.9 took place. Similarly, during the C7.0 flare that we analyzed [\texttt{SOL2014-11-07T16:10}], first the $\delta$-complex \textit{A} was involved, with indication of a filament motion, and successively the $\delta$-complex \textit{B} hosted the circular ribbon of the flare. Also during this event flare-like brightenings moved southward from the $\delta$-complex \textit{A} to the $\delta$-complex \textit{B} (see Figure~\ref{fig:flareC}, 16:25~UT). 

The X1.6 flare was analyzed by \citet{yurchy15}, who reported that this flare began as a localized eruption of core fields inside the northern $\delta$-complex \textit{A}, triggered by flux emergence occurring at the boundary between the two umbrae. Later, the event involved the entire AR, releasing a fast and wide coronal mass ejection (CME). They found that this event was accompanied by post-eruption arcades, J-shaped flare ribbons exhibiting fine structures, and irreversible changes in the magnetic configuration in the photosphere. 

According to \citet{yurchy15}, the X1.6 flare and the related filament eruption were triggered by magnetic flux emergence at the boundary between the two umbrae of opposite polarity of the $\delta$-complex \textit{A}. To support this claim, they rely on the analysis of the magnetic flux trend within these umbrae using HMI measurements (see Figure~1 in \citealp{yurchy15}). However, while the authors indicate that they employed HMI data from the \textit{hmi.B\textunderscore720s} series, which of course have a cadence of 12~minutes, in the plot the cadence appears to be different. 

Our measurements of the total magnetic flux in that area disagree with those reported by \citealp{yurchy15}: the total magnetic flux in the area remains almost constant, rather showing a slight decrement (Figure~\ref{fig:flux}). There is only an increase of the negative magnetic flux, which is roughly compensated by the decrease of the positive magnetic flux. Indeed, Figure~\ref{fig:magnetic} clearly shows that in the interface region between the two opposite polarity umbral cores of the $\delta$-complex \textit{A}, the magnetic field changes sign (fourth panel). In addition, there is a variation of distribution of the total field strength toward the north of the $\delta$-complex \textit{A}: the magnetic field strengthens in the left area and decreases in the right area, mainly due to an analogue variation in the strength of the horizontal field.

Therefore, from our analysis it appears that flux emergence cannot be invoked as the trigger mechanism of these flares. In contrast, combining our findings with the analysis carried out by \citet{yurchy15}, it can be deduced that the eruption of a flux rope, occurred in the $\delta$-complex \textit{A} region, where the presence of strongly sheared field lines was witnessed by penumbral filaments aligned along the PIL between the two opposite umbral cores (see Figure~5 in \citealp{yurchy15}). After the C7.0 and X1.6 flares, penumbral filaments reached a more relaxed configuration, as demonstrated by the azimuth changes detected by \citet{yurchy15}. The increasing energy content of the flares occurred before the onset of the X1.6 flare, together with the precursor brightenings, the shaking motion of the filament connecting the $\delta$-complexes \textit{A} and \textit{B} seen during the C7.0 flare, and its sigmoidal configuration along the PIL, just before the X1.6 flare, suggest that the AR~12205 was progressively destabilized. In this perspective, the eruption of the flux rope observed at the beginning of the X1.6 flare seems to have triggered a kind of domino effect \citep[see, e.g.,][]{Zuccarello:09}, which eventually caused the launching of the filament settled along the PIL that later formed the CME observed by LASCO.

\subsection{Emission in the continuum and line profiles}

During the rise phase of the X1.6 flare (at 16:57~UT), signatures of emission in the continuum close to the wings of the \ion{Mg}{2}~k line were detected (\textit{IRIS} dataset), while a few minutes before the flare peak, a ribbon was observed in the \textit{G} band and in the 4170~\AA{} continuum (ROSA dataset). The areas of enhanced emission involved $\delta$ sunspots in the northern and southern parts of AR~12205. In particular, around the peak time of the X1.6 class flare, the ROSA dataset shows the presence of bright ribbons at continuum wavelengths separating at an average velocity of $\approx 10 \,\mathrm{km\,s}^{-1}$. 

Concerning the physical processes at work, it has been suggested that hydrogen recombination in the chromosphere might have an important role in flare WL emission. More specifically, the hydrogen recombination occurring in the chromosphere may be related to the so-called radiative back-warming, so that there is an increase of temperature in the photosphere that can produce the optically thick flare emission in the continuum \citep{hudson:72,Metcalf:90a,Metcalf:90b,Ding:96}.

With regard to UV line profiles, it is generally believed that the transport of energy from the coronal reconnection site to the chromosphere implies the occurrence of a beam of accelerated (non-thermal) electrons that impact the chromosphere and cause chromospheric evaporation. Observations indicate the presence of chromospheric heating such that hot ($8 - 25$~MK) upflows (up to $-400$~\kms{}) along the flare loops are detected in SXR and EUV. The heated plasma expands upward at the sound speed, while at the same time, the chromospheric plasma is strongly compressed and downward propagating shock waves are excited in the chromosphere.

During the impulsive phase of flares, some authors observed redshifts and enhancements in the red wing of chromospheric lines \citep[][and references therein]{Tei:18}. In H$\alpha$ downward plasma velocities of $\sim 50$~\kms{} have been measured and have been interpreted in terms of momentum balance with upflows of the chromospheric plasma \citep{Ichimoto:84}.

\citet{Svestka:62} analyzed the line asymmetries in the spectra of 92 flares and found evidence that 80\% of them showed red asymmetry and 23\% shows blue asymmetry. Later, \citet{Heinzel:94} found that the Balmer and the \ion{Ca}{2}~H lines show blue asymmetry during the onset phase of flares.

\citet{Kerr:15} using \ion{Mg}{2} \textit{IRIS} data found redshifts equivalent to velocities of $15 - 26$~\kms{}, while \citet{Graham:15} found upflows of up to $300$~\kms{} and downflows up to $40$~\kms{} during the impulsive phase.

\citet{Tei:18} analyzed a C-class flare using data in the \ion{Si}{4} 1403~\AA, \ion{C}{2} 1335~\AA, and \ion{Mg}{2}~h\&k lines from \textit{IRIS} and the \ion{Ca}{2}~K, \ion{Ca}{2} 8542~\AA, and H$\alpha$ lines from the Domeless Solar Telescope. They found that in the \ion{Mg}{2}~h line, the leading edge of the flare kernel showed an intensity enhancement in the blue wing, and a smaller intensity of the blue-side peak than that of the red-side. The blueshift lasted for $9 - 48$~s with a typical speed of $10.1 \pm 2.6$~\kms{} and it was followed by the high intensity and the large redshift with a speed of up to $50$~\kms{} detected in the \ion{Mg}{2}~h line. The large redshift was a common property for all six lines but the blueshift prior to it was found only in the \ion{Mg}{2} lines. Similarly, studying \ion{Mg}{2} lines during an M6.5 flare, \citet{Huang:19} observed blue-wing enhancement, with typical blueshifts of about $10$~\kms{}, up to $20$~\kms{}, and strong broadening on the leading edge of the propagating ribbon, while redshifts in the trailing areas. Comparing their observations with numerical modeling, \citet{Huang:19} suggested that the enhanced blue wings in \ion{Mg}{2} lines can be due to an increase of local electron density and a decrease in temperature, caused by electron precipitation, whereas a spatially unresolved turbulence of $10 - 30$~\kms{} can be responsible for the broadening.

The analysis carried out in this work shows that for \ion{Si}{4} (see Figure~\ref{fig:iris_profiles}, top panel), the line profiles and the velocity values for the pixel positions indicated by the orange and blue circles in Figure~\ref{fig:iris_zoom} indicate upflows followed by downflows, while the line profile at the position denoted by the green circle is indicative of downflows in the analyzed time interval.

If we interpret these upflows and downflows in the framework of processes of evaporation and condensation associated to the standard flare model, we can conclude that in two among the three pixels examined, a process of chromospheric evaporation is followed by a condensation, while in the third slit position there is continuous condensation.

In the framework of the same scenario, for \ion{C}{2} (see Figure~\ref{fig:iris_profiles}, second panel) in two among the three pixel positions examined, a process of chromospheric evaporation is followed by a condensation (orange and blue profiles), while in the third slit position there is continuous condensation (green profiles).

In the case of \ion{Mg}{2}~k\&h (see Figure~\ref{fig:iris_profiles}, third and fourth panels), the line profiles and the velocity values for the pixel positions indicated by the orange and blue circles in Figure~\ref{fig:iris_zoom} demonstrate the presence of upflows between 16:54 and 16:56 UT, while the line profile at the position indicated by the green circle suggests downflows. Different pixel positions indicate that at the same time, process of chromospheric evaporation and condensation are taking place. 

In summary, in this scenario, for selected time intervals and slit positions at the flare ribbon, the line profiles of \ion{C}{2} 1335.75~\AA{}, \ion{Si}{4} 1402.8~\AA{}, and \ion{Mg}{2}~k\&h lines may suggest the occurrence of a process of chromospheric evaporation followed by a condensation taking place at different atmospheric heights, provided the different formation temperatures of the lines. The rising flux rope may explain an electron influx to the ribbon, causing the evaporation as well as providing a source for the continuum enhancements.

An alternative explanation for the presence of blueshifts at some slit positions could be related with the upward motion of the erupting filament at the beginning of the X1.6 flare. Recently, \citet{Kleint:15} found Doppler shifts ranging between $-100$ to $-600$~\kms{} in \textit{IRIS} \ion{Si}{4} spectra during the filament eruption leading to an X1 flare. Typical velocities of eruptions of filaments are of the same order of magnitude, ranging from less than $100$~\kms{} up to $1000$~\kms{}, even if they depend on the height of the measurements \citep[see][and references therein]{Kleint:15}. 

The upward velocities inferred in our study from the analysis of different lines range between $-50$~\kms{} to $-100$~\kms{}, therefore we cannot completely discard the hypothesis that these blueshifts might be related to the motions of the uprising filament. However, our analysis of \textit{IRIS} data relevant to the flux rope shows a rather different pattern in the spectroheliograms and in the individual line profiles, the upward velocities of the erupting filament being in generally slightly larger than those observed in the ribbon.

On the other hand, an enhancement of continuum emission in FUV and NUV, as well as a very prominent bump in the blue wing of the \ion{Si}{4}, \ion{C}{2}, and \ion{Mg}{2}~h\&k lines were detected (see Figure~\ref{fig:iris_continua}). Based on the previous considerations, we propose that this bump cannot be attributed to the upward velocity component of the ejected flux rope, since the line profiles in this structure do not exhibit such a broadened line. Indeed, the \textit{IRIS} slit caught the structure during the launch phase, as shown in Figure~\ref{fig:iris_evolution}. The bump is more prominent higher in the atmosphere (\ion{Si}{4}) than at chromospheric heights (\ion{Mg}{2}~h\&k lines). This could suggest that we see the effect of the reconnection due to magnetic braiding, as during the launch of jets \citep[see, e.g,][]{Huang:18}, resulting in a large, non-thermal broadening of the lines.

\subsection{Changes in the penumbrae}

With regard to the observed changes in the penumbrae around the sunspots involved in the C7.0 and X1.6 flares, our analysis highlights that these are due to magnetic fields in the regions around sunspots becoming more
vertical or horizontal, leading to the decay or enhancement of penumbral regions, respectively. This finding confirms the suggestion by \citet{Liu:05} about the flare-associated changes in white-light continuum intensity being related with permanent variations in the inclination as a result of the reconnection in $\delta$ sunspots. It also supports the close correlation between the changes in the sunspot intensity and horizontal field strength which was studied by \citet{Song:16} in a sample of flaring ARs, including AR~12205 after the X1.6 flare. 

Recent observations of the magnetic field rearrangement in the photosphere after an M5.0 solar flare show that the inner penumbral enhancement around the flaring PIL area was accompanied by the field collapsing down, whereas the outer penumbral decay area was associated with the field lifting up toward the upper flare center \citep{Xu:19}. Contrarily, 3D simulations suggest that observed enhancement in the photospheric horizontal magnetic fields along the PIL results from the reconnection-driven contraction of sheared flare-loops, which increases the downward component of the Lorentz force density around the PIL \citep{Barczynski:19}.

AR~12205 was also included in the sample of ARs investigated by \citet{Lu:19} to evaluate the magnetic imprints of X-class flares in the photosphere, in the context of back reaction on the solar surface associated to coronal field restructuring \citep{Hudson:08,WangLiu:10}. In this study, the X1.6 flare of AR~12205 was considered an event that increased the horizontal field component only along the PIL \citep{Lu:19}.

In our observations we also found evidence for penumbral formation (see the region at around [$40\arcsec, 40\arcsec$] in Panel 8, Figure~\ref{fig:sot}). \citet{yurchy15} observed that these penumbral-features, similar to orphan penumbrae \citep[e.g.][]{Lim:13}, were developed under a stable filament, being not related to the presence of opposite-polarity fields around a PIL. We suggest that the overlying filament is trapping part of the magnetic flux whose orientation has been modified by the flare into a more horizontal configuration. This is supported by the fact that we did not observe any variation of the magnetic flux in the \textit{Hinode} sub-FOV around $\delta$-complex \textit{A} (see Figure~\ref{fig:flux}, top panel), therefore it appears that the formation of these penumbral-like structures is not linked to flux emergence. 
Indeed, invoking such a mechanism able to lead to the presence of highly inclined, nearly horizontal magnetic fields, for being responsible for penumbra formation (or, conversely, decay) strengthens the argument which has been proposed in recent studies concerning the development of penumbrae and penumbral-like structures, from both the observational \citep[see, e.g.][]{Shimizu:12,Lim:13,Guglielmino:14,Romano:13,Romano:14,Zuccarello:14,Jurcak:14,Jurcak:17,Murabito:17,Murabito:18} and the theoretical point of view \citep[][]{Rempel:12,MacTaggart:16}, even in absence of magnetic flux variations. In this perspective, our findings indicate that the effects of magnetic reconfiguration leading to penumbral enhancements driven by flares can extend well beyond the region around the flaring PILs.

In connection with back reaction analysis, our findings relevant to $\delta$-complex \textit{A} are in agreement with the qualitative observational signature proposed by \citet{WangLiu:10}: the observed limbward flux increases while diskward flux decreases rapidly and irreversibly after the C7.0 and X1.6 flare. This can be easily noticed in the difference image between the final and initial map of vertical field component (Figure~\ref{fig:magnetic}, panel four) and in the graph of the relative magnetic flux in Figure~\ref{fig:flux} (bottom panel).

\section{Conclusions}

We studied two consecutive C7.0 and X1.6 flares occurred in AR~12205 using data acquired by ground-based and satellite instruments. Our analysis brought new information to interpret these events, which showed a complex behaviour.

Both the \textit{IRIS} and the ROSA datasets show the presence of bright ribbons at continuum wavelengths around the peak time of the X1.6 class flare. The WL emission, if interpreted in the framework of hydrogen recombination in the chromosphere, may be related to radiative back-warming due to an increase of temperature in the photosphere.

The interpretation of the line profiles of \ion{C}{2} 1335.75~\AA{}, \ion{Si}{4} 1402.8~\AA{}, and \ion{Mg}{2}~k\&h lines has been discussed in two scenarios. The former, based on the hypothesis of the occurrence of plasma motions foreseen in the standard flare model, seems to suggest the occurrence of a process of chromospheric evaporation followed by condensation, which takes place at different atmospheric heights, as seen in lines with different formation temperatures. The latter scenario, which seems to be the most plausible, taking into account both the line profiles and the presence of a very prominent bump in the blue wing of the \ion{Si}{4}, \ion{C}{2}, and \ion{Mg}{2}~h\&k, is based on the hypothesis that the blueshifts detected at some slit positions are actually indicative of the rising motion of an eruptive filament. 

As far as the flare's triggering mechanism is concerned, comparing our analysis with the one carried out by \citet{yurchy15}, we can conclude that the eruption of a flux rope triggered these events, so that the release of the shear stored in the field lines, rather than flux emergence, is the main reason for the flare occurrence.

In this scenario, tether-cutting reconnection could be considered a plausible mechanism for the formation of the unstable flux rope \citep{Moore:01,yurchy06,Xue:17,Chen:18}. Such a mechanism is often invoked to explain the back reaction on the solar surface that we have noticed in our observations as well \citep[e.g.,][]{WangLiu:10,Lu:19}. On the other hand, slipping reconnection seems to be at work during these events, as witnessed by the appearance of the bright elongated patch observed to the east of the $\delta$-spot region a few minutes after the onset of the X1.6 flare. Indeed, such a configuration with a third, remote brightening site is reminiscent of a fan-spine topology, with the dome-shaped fan located above the $\delta$-complex \textit{A} \citep[see, e.g.,][]{Guglielmino:16}. This seems to be confirmed by the analysis of the previous C7.0 flare.

Finally, concerning the changes in the penumbrae observed during the flares, our analysis provided indications that the overlying filament can trap part of the magnetic flux system that has been modified by the flare, becoming more horizontal, and that the magnetic reconfiguration can take place also in a region far from the PIL.

We believe that the next generation of ground-based telescopes, like the European Solar Telescope \citep{Collados:10} and the Daniel K.~Inouye Solar Telescope \citep{Keil:10} will be fundamental to shed light on some of the issues that remain to be further investigated and clarified. These issues include (but are not limited to) the possibility to impute, and to what degree, how flare triggering is related to magnetic flux emergence and/or shearing and how to distinguish the motion associated with an erupting filament with processes associated to plasma evaporation/condensation. Another key process that remains to be addressed is the relationship between continuum enhancement and magnetic field rearrangement.

\acknowledgments

The authors wish to thank an anonymous referee for his/her helpful comments. The research leading to these results has received funding from the European Commissions Seventh Framework Programme under the grant agreements no.~606862 (F-CHROMA project), and no.~312495 (SOLARNET project) and from the European Union's Horizon 2020 research and innovation programme under grant agreement no.~739500 (PRE-EST project) and no.~824135 (SOLARNET project). This work was supported by the Italian MIUR-PRIN grant 2012P2HRCR on The active Sun and its effects on space and Earth climate, by Space Weather Italian COmmunity (SWICO) Research Program, and by the Universit\`a degli Studi di Catania (Piano per la Ricerca Universit\`{a} di Catania 2016-2018 -- Linea di intervento~1 ``Chance''; Linea di intervento~2 ``Ricerca di Ateneo - Piano per la Ricerca 2016/2018''). P.H.K. is grateful to the Leverhulme Trust for the award of an Early Career Fellowship. \textit{IRIS} is a NASA small explorer mission developed and operated by LMSAL with mission operations executed at NASA Ames Research center and major contributions to downlink communications funded by ESA and the Norwegian Space Centre. The \textit{SDO}/HMI data used in this paper are courtesy of NASA/\textit{SDO} and the HMI science team. \textit{Hinode} is a Japanese mission developed and launched by ISAS/JAXA, with NAOJ as domestic partner and NASA and STFC (UK) as international partners. It is operated by these agencies in co-operation with ESA and Norwegian Space Centre. Use of NASA's Astrophysical Data System is gratefully acknowledged.

\facility{\textit{Hinode} (SOT), \textit{IRIS}, \textit{SDO} (HMI, AIA)}

\end{document}